\documentclass[12pt]{article}

\pdfoutput=1

\usepackage{amsmath}
\usepackage{bbm}

\usepackage{amssymb}
\usepackage{graphicx}
\newcommand{\Sec}[1]{Sec.~\ref{#1}}
\newcommand{\Secs}[2]{Secs.~\ref{#1} and \ref{#2}}
\newcommand{\App}[1]{App.~\ref{#1}}
\newcommand{\Tab}[1]{Table~\ref{#1}}
\newcommand{\Tabs}[2]{Tables~\ref{#1} and \ref{#2}}
\newcommand{\Fig}[1]{Fig.~\ref{#1}}
\newcommand{\Figs}[2]{Figs.~\ref{#1} and \ref{#2}}
\newcommand{\Eq}[1]{Eq.~(\ref{#1})}
\newcommand{\Eqs}[2]{Eqs.~(\ref{#1}) and (\ref{#2})}
\newcommand{\Ref}[1]{Ref.~\cite{#1}}
\newcommand{\Refs}[1]{Refs.~\cite{#1}}

\newcommand{\be}{\begin{equation}}
\newcommand{\ee}{\end{equation}}
\newcommand{\bea}{\begin{eqnarray}}
\newcommand{\eea}{\end{eqnarray}}

\begin{document}
\baselineskip 0.6cm

\def\simgt{\mathrel{\lower2.5pt\vbox{\lineskip=0pt\baselineskip=0pt
           \hbox{$>$}\hbox{$\sim$}}}}
\def\simlt{\mathrel{\lower2.5pt\vbox{\lineskip=0pt\baselineskip=0pt
           \hbox{$<$}\hbox{$\sim$}}}}
\def\one{\relax{\rm 1\kern-.25em 1}}

\begin{titlepage}

\begin{flushright}
MIT-CTP 4156
\end{flushright}

\vskip 1.0cm

\begin{center}

{\Large \bf 
The Bestest Little Higgs
}

\vskip 0.6cm

{\large
Martin Schmaltz$^1$, Daniel Stolarski$^{2,3}$, and Jesse Thaler$^4$
}

\vskip 0.4cm

{\it $^1$ Physics Department, Boston University, Boston, MA 02215} \\
{\it $^2$ Berkeley Center for Theoretical Physics, Department of Physics,\\
  University of California, Berkeley, CA 94720} \\
{\it $^3$ Theoretical Physics Group, Lawrence Berkeley National 
  Laboratory, Berkeley, CA 94720} \\
{\it $^4$ Center for Theoretical Physics, 
  Massachusetts Institute of Technology, Cambridge, MA 02139}

\vskip 0.8cm

\abstract{While little Higgs models provide an interesting way to address the hierarchy problem, concrete models in the literature typically face two major obstacles.  First, the mechanism for generating a Higgs quartic coupling often leads to large violations of custodial symmetry.  Second, there is a tension between precision electroweak observables in the gauge sector and fine-tuning in the top sector.  In this work, we present a new little Higgs model which solves both of these problems.  The model is based on an $SO(6)\times SO(6)/SO(6)$ coset space which has custodial symmetry built in.  The Higgs quartic coupling takes a particularly simple form and does not suffer from the ``dangerous singlet'' pathology.  We introduce a gauge breaking module which decouples the mass of gauge partners from the mass of top partners, allowing for natural electroweak symmetry breaking.  The collider phenomenology is dominated by production and decay of the top partners, which are considerably lighter than in traditional little Higgs theories.}

\end{center}
\end{titlepage}

\setcounter{tocdepth}{1}

\tableofcontents

\section{Introduction}
\label{sec:intro}

While the Standard Model (SM) has been tremendously successful in predicting the results of experiments, it suffers from the hierarchy problem.  Namely, quantum corrections render the Higgs potential fine-tuned.  These quantum corrections come from three different sectors of the SM: the gauge sector, the fermion sector, and the Higgs sector.  In order to have a theory which is not fine-tuned, new physics is needed in each of these three sectors to cancel quantum corrections to the Higgs potential.

Little Higgs theories~\cite{ArkaniHamed:2001nc, ArkaniHamed:2002qx, ArkaniHamed:2002qy, Schmaltz:2005ky} address the hierarchy problem through so-called ``collective symmetry breaking'', which is implemented for the gauge, fermion, and Higgs quartic couplings.   In these models, the Higgs is a pseudo-Nambu--Goldstone boson (PNGB), like in the original composite Higgs scenarios~\cite{Kaplan:1983sm, Dugan:1984hq}.  Below a ``compositeness'' scale $\Lambda \sim 10~\text{TeV}$, the Higgs potential is protected by global shift symmetries, which are then broken collectively.  That is, the symmetries are broken by two different operators, each of which leaves a subgroup of the symmetry unbroken so that neither operator alone can generate a Higgs potential.  The combination of the two operators breaks enough of the symmetries to generate a Higgs potential, but a quadratically-divergent Higgs mass is not generated at one loop.  Collective symmetry breaking implies the existence of partner particles for most SM fields, leading to rich collider phenomenology~\cite{Perelstein:2005ka}.  

While collective symmetry breaking does control quantum corrections to the Higgs potential, concrete little Higgs models generically face two major obstacles:
\begin{enumerate}
\item \textbf{Custodial symmetry and the Higgs quartic coupling.}  While it is relatively straightforward to implement collective symmetry breaking in the gauge and fermion sectors, it is more difficult to generate a collective Higgs quartic coupling.  As shown in Ref.~\cite{Schmaltz:2008vd}, a collective quartic requires extra PNGBs with specific electroweak quantum numbers, which already imposes some model building constraints.  Adding to the challenge, the full PNGB sector must preserve a custodial $SU(2)$ symmetry in order to avoid excessive contributions to the $T$-parameter from the PNGB kinetic terms. 

\item \textbf{Gauge partners vs.\ fermion partners.}  Precision electroweak measurements strongly constrain new physics which can mix with SM gauge bosons~\cite{Peskin:1990zt, Peskin:1991sw}.  For little Higgs models without $T$-parity~\cite{Cheng:2003ju,Cheng:2004yc,Low:2004xc}, this means that the gauge boson partners must be rather heavy \cite{Csaki:2002qg, Csaki:2003si}, $m_{W'} \simgt 2\!-\!3$ TeV. On the other hand, avoiding fine-tuning in the top sector imposes an upper bound on the mass of the top partner, $m_T \simlt 1\!-\!2$ TeV. Thus avoiding fine-tuning requires light top partners whereas precision electroweak constraints require heavy gauge boson partners. Unfortunately, most little Higgs models predict the opposite generic relation
\be
\frac{m_T}{m_{W'}} \simeq \frac{m_t}{m_W} \simeq 2  .
\label{eq:badrelation}
\ee
\end{enumerate}

Previous attempts at solving these problems have not been completely successful. In \Ref{Chang:2003un} a Higgs potential with collective symmetry breaking and custodial symmetry was constructed, however custodial $SU(2)$ was violated by the Higgs vacuum expectation values (vevs). The model in \Ref{Chang:2003zn} does have custodial $SU(2)$, but it suffers from quadratic divergences due to a ``dangerous singlet"~\cite{Schmaltz:2008vd} in the quartic potential.  The troublesome relationship in \Eq{eq:badrelation} is avoided in ``Simple Group" models~\cite{Kaplan:2003uc,Schmaltz:2004de,Skiba:2003yf} in a region of parameter space where gauge boson partners and top partners obtain masses proportional to different symmetry breaking scales. However, even in the preferred region of parameter space~\cite{Han:2005dz} the top partner still ends up being heavier than the gauge boson partners.

In this paper, we construct a little Higgs model which solves both of these problems. Custodial $SU(2)$ and relatively heavy gauge boson partners allow the model to evade constraints from precision electroweak measurements, and fine-tuning is avoided because the top partners are quite light. 

The new ingredient which allows us to succeed is a modular approach to model building.  To start, we build a non-linear sigma model with a collective Higgs quartic coupling and custodial $SU(2)$.  This model has the global symmetry breaking pattern $SO(6)\times SO(6)/SO(6)$ and a decay constant $f$. It can be represented by a moose diagram \cite{Georgi:1985hf} with two sites and only one link field, making it far simpler in structure than most little Higgs models.   A diagonal unbroken $SO(4)$ subgroup is identified with the $SU(2)_L \times SU(2)_R$ of the SM, guaranteeing that the theory has an approximate custodial $SU(2)$ symmetry.  The PNGBs include two Higgs doublets, which get a collective quartic coupling with the help of an electroweak singlet PNGB.  Because this singlet is not ``dangerous''~\cite{Schmaltz:2008vd}, the quartic coupling is viable.  It is then straightforward to introduce SM fermions and generate a collective top Yukawa coupling $\lambda_t$.  We implement the top sector in a way that yields a particularly light top partner, with the mass of the top partner of order $(2/3) \lambda_t f$.\footnote{This factor of $2/3$ should be compared to the factor of $2$ often found in little Higgs constructions, leading to a corresponding decrease in fine-tuning from the top sector by a factor of 9, for fixed decay constant $f$.} 

Armed with a viable Higgs (and fermion) sector, we want to control quantum corrections from SM gauge loops.    If we were to independently gauge $SU(2)\times U(1)$ subgroups of each global $SO(6)$ symmetry, then the quadratic divergences would be canceled, but the $SO(6)\times SO(6)$ symmetry breaking would give the gauge boson partners masses of order $g_{\rm EW} f$, i.e.\ lighter than the top partner.   To avoid this problem we proceed in a modular way.  We introduce a second non-linear sigma field with a decay constant $F$ which is a singlet under the global symmetries of the Higgs and top sectors, but which transforms under the same gauge symmetries as the Higgs sector.   When this second non-linear sigma field gets a vev, it breaks the gauge group down to the SM group, giving masses to the gauge partners of order $g_{\rm EW}F$.  Crucially, because $F$ can be made larger than $f$, we can raise the mass of the gauge partners without affecting the mass of the top partners.  Thus, the theory can be made safe from electroweak precision constraints without introducing large fine-tuning from the top sector.  This technique is quite general, and can be implemented in almost any little Higgs model with a viable collective quartic.

Like ordinary little Higgs models, our construction predicts top quark partners and their corresponding collider signatures.  However, in our case, the top partners are particularly light, on the order of 500--800 GeV compared to the $\sim2$ TeV top partners predicted in many little Higgs models without $T$-parity.  Thus, the collider phenomenology is dominated by production and decay of the top partners, yielding a final state consisting of two third generation fermions and between two and four gauge or Higgs bosons.  In addition, our modular collective gauge sector necessarily implies the existence of light PNGBs that dominantly couple to third generation fermions and transform as a $\mathbf{6}$ of $SO(4)$, namely an $SU(2)_L$ triplet, a complex singlet with hypercharge, and a real neutral singlet.\footnote{These light PNGBs can be roughly thought of as the would-be longitudinal components of the heavy gauge bosons in an ordinary little Higgs theory.  In our construction, these extra PNGBs are left uneaten, and since they couple mainly to the third generation, they face no precision electroweak bounds.}  Discovery of these uneaten PNGBs modes---and the discovery of relatively heavy gauge boson partners---would be strong evidence for the scenario proposed here.

The organization of this paper is as follows.  In \Sec{sec:higgs}, we describe the non-linear sigma model which contains the Higgs sector and explain how we generate a quartic coupling for the Higgs.  We also describe the Higgs potential and show that we get satisfactory electroweak symmetry breaking.  In \Sec{sec:gauge}, we describe our modular approach to implementing a collective gauge coupling and show that custodial symmetry is indeed preserved.  We also describe how to implement hypercharge.  In \Sec{sec:fermion}, we describe the fermion sector of the theory including the collective top Yukawa coupling.  In \Sec{sec:pew}, we detail the constraints on the model, particularly from precision electroweak measurements, and show the allowed regions of parameter space.  A brief description of the collider phenomenology of our model is given in \Sec{sec:collider}, and conclusions are given in \Sec{sec:conc}.  Various calculational details and alternative possibilities are left to the appendices.  

\section{Higgs Sector}
\label{sec:higgs}

The lesson from Ref.~\cite{Schmaltz:2008vd} is that a collective quartic coupling in little Higgs theories requires additional PNGBs with specific quantum numbers:  a one-Higgs doublet model requires an electroweak triplet, and a two-Higgs doublet model requires either an electroweak triplet or singlet.  Without $T$-parity~\cite{Cheng:2003ju,Cheng:2004yc,Low:2004xc}, an electroweak triplet will typically get a vev and violate custodial $SU(2)$, so our starting point will be a two-Higgs doublet model with an additional singlet and manifest custodial $SU(2)$ symmetry.

The simplest symmetry breaking pattern that would yield these PNGBs is $SO(6)/SO(4)$, but this coset space is cumbersome, both for getting a collective quartic coupling and for implementing the modular collective gauge coupling we need in \Sec{sec:gauge}.  Recently, \Ref{Hook:2009kx} classified all the simple group coset spaces which admit a collective quartic without dangerous singlets.  We instead take a different route, inspired by Ref.~\cite{Thaler:2005kr}, and start with a product group coset space $SO(6)_A \times SO(6)_B / SO(6)_V$, which contains the desired PNGBs, as well as additional PNGBs that will somewhat affect the collider phenomenology of this model.  

\subsection{Non-Linear Sigma Structure}
\label{sec:nlsigma}

Under the global $SO(6)_A \times SO(6)_B$ symmetry, we introduce a non-linear sigma field that transforms as:
\begin{equation}
\Sigma \rightarrow G_A \Sigma\, G_B^\dagger \ .
\end{equation}
All the group elements of $SO(6)$ are real so $G^\dagger = G^T$, and $\Sigma$ is in a real representation of $SO(6)\times SO(6)$ so $\Sigma^\dagger = \Sigma^T$.  The vev of $\Sigma$,
\be
\langle \Sigma \rangle = \one \ ,
\ee
spontaneously breaks the global symmetry down to the diagonal $SO(6)_V$.  

\begin{figure}
\center{\includegraphics[scale=0.8]{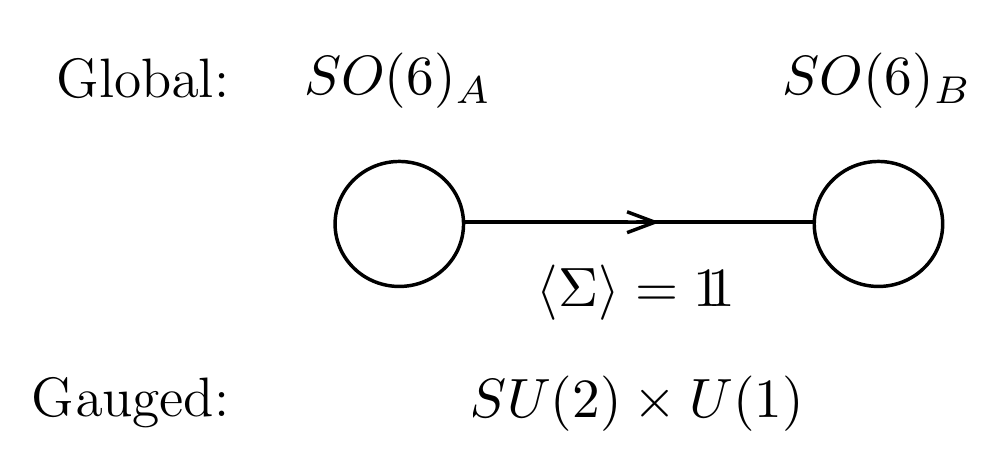}}
\caption{Moose diagram for the simple model in \Sec{sec:higgs}.  There is a global $SO(6)_A \times SO(6)_B$ symmetry with a diagonal $SU(2)\times U(1)$ gauged.  $\Sigma$ transforms as a bifundamental of the global symmetry, and gets a vev which spontaneously breaks the global symmetry down to $SO(6)_V$ and preserves the gauge symmetry.}
\label{fig:moosediagram}
\end{figure}

The upper $SO(4)$ block in each $SO(6)$ can be decomposed into $SU(2)_L \times SU(2)_R$, with the corresponding six generators $T_L^a$ and $T_R^a$ given in \App{sec:grouptheory}.  We weakly gauge the diagonal combination of $SU(2)_{LA}$ and $SU(2)_{LB}$ and identify this with the SM $SU(2)_{L}$ gauge group.  We also gauge the diagonal combination of the third component of $SU(2)_{R}$ ($T^3_{RA} + T^3_{RB}$) and identify it with SM hypercharge.   The symmetry structure of the theory is exhibited in ``moose''  \cite{Georgi:1985hf} notation in \Fig{fig:moosediagram}.  

Note that the gauge structure described so far explicitly breaks the symmetries protecting the Higgs. It is not collective and will lead to quadratically-divergent contributions to the Higgs mass. This is what we intended, as we want to make the gauge boson partners parametrically heavier than the decay constant $f$ in the $\Sigma$ sector in order to avoid precision electroweak constraints.   The gauge boson quadratic divergences will be canceled at a higher scale $F > f$ where a new gauge structure emerges, and the details of this are described in \Sec{sec:gauge}.  

The PNGBs from $SO(6)_A \times SO(6)_B / SO(6)_V$ can be parameterized by looking at broken symmetry transformations about the vev of $\Sigma$
\be
\Sigma = G \langle \Sigma \rangle G = G^2  \langle \Sigma \rangle =
e^{2 i \Pi^a T^a/f} \ .
\label{eq:param1}
\ee
Here, $f$ is the decay constant which we take to be of order a TeV.   We will be agnostic as to what high energy dynamics gives rise to this non-linear sigma model, so our construction will require some kind of UV completion above the ``compositeness''  scale $\Lambda \sim 4 \pi f$.\footnote{In general, unitarity could break down at a scale lower than $4\pi f$ \cite{Chang:2003vs,Espinosa:2006nu}, but the question of precisely what states unitarize the $\Sigma$ model is beyond the scope of this paper.}   If we were using an $SU(N)$ group, then the normalization of the generators would be ${\rm Tr}(T^a T^b) = \frac{1}{2} \delta^{ab}$.  Because the algebra of $SO(6)$ is isomorphic to that of $SU(4)$, we can construct the correspondence between the structure constants which requires ${\rm Tr}(T^a T^b) = \delta^{ab}$ for $SO(6)$.  The correspondence between the algebras is given in \App{sec:grouptheory}.  

Instead of the parameterization of \Eq{eq:param1}, we take the following more convenient parameterization
\begin{equation}
\Sigma = e^{i \Pi/f}  e^{2i \Pi_h/f}  e^{i \Pi/f}  .
\label{eq:sig}
\end{equation}
The matrices $\Pi$ and $\Pi_h$ are $6\times6$ imaginary anti-symmetric matrices given by:  
\begin{equation}
\Pi =  i
  \left( \begin{array}{ccc}
   \phi_a T_L^a + \eta_a \, T_R^a & 0 & 0 \\
    0 & 0 & \sigma/\sqrt{2}  \\
    0 & -\sigma/\sqrt{2}  & 0
  \end{array} \right),
  \label{eq:pi}
\end{equation}
\begin{equation}
\Pi_h =   \frac{i}{\sqrt{2}}
  \left( \begin{array}{ccc}
    0_4 & h_1 & h_2 \\
    - h_1^T & 0 & 0 \\
    - h_2^T & 0 & 0
  \end{array} \right),
  \label{eq:pih}
\end{equation}
where the first row and column represent vectors of length four. The $h_i$ are $\mathbf{4}$'s of $SO(4)$ which can also be thought of as complex $SU(2)_L$ doublets with hypercharge and will be identified as the scalars of a two Higgs doublet model.  The correspondence between the two types of notation is described further in \App{sec:grouptheory}.  The $\sigma$ is a real singlet field that will be crucial for getting a collective quartic coupling.  The $T_{L,R}^a$ are the generators of the two $SU(2)$s contained in $SO(4)$ and are given in \App{sec:grouptheory}.  The $\phi_a$ form an electroweak triplet with zero hypercharge.  The $\eta_a$ transform as a triplet under $SU(2)_R$, that is, the $\eta_1$ and $\eta_2$ components form a complex singlet with hypercharge, and $\eta_3$ is a real singlet.  Because we have only gauged unbroken diagonal generators in $SO(4)_A \times SO(4)_B$, none of the $\phi_a$ or $\eta_a$ PNGBs are eaten.  

We choose the parameterization in \Eq{eq:sig} for several reasons. We separated out the Higgs fields from the other PNGBs in order to avoid mixing between them once electroweak symmetry is broken. In particular, in \Sec{sec:collectivequartic} we will see that the fields in the upper  $4\times4$ block of $\Pi$ will not appear in the Higgs quartic sector.  Finally, this parameterization preserves a pleasing $\Sigma \leftrightarrow \Sigma^T$ symmetry which acts as $\Pi \rightarrow -\Pi$ and $\Pi_h \rightarrow -\Pi_h$ on the PNGBs. 

\subsection{Collective Quartic Coupling}
\label{sec:collectivequartic}

In order to generate a viable Higgs quartic coupling, we must explicitly break some of the symmetries under which the Higgses transform non-linearly.  Collective symmetry breaking requires two operators, each of which explicitly breaks some of the global symmetries, but neither by itself would allow the Higgs to get a potential.  In order to do this, we define the following projectors:
\begin{equation}
P_5 = \mbox{diag}(0,0,0,0,1,0) \qquad P_6 = \mbox{diag}(0,0,0,0,0,1) \ .
\label{eq:P56}
\end{equation}
The collective quartic potential is then given by
\begin{equation}
V_{\rm quartic} = 
    \frac{1}{4} \lambda_{65} f^4\, {\rm tr} \left( P_6 \Sigma P_5 \Sigma^{T} \right)
    + \frac{1}{4} \lambda_{56} f^4\, {\rm tr} \left( P_5 \Sigma P_6 \Sigma^{T}  \right)
    = \frac{1}{4} \lambda_{65} f^4 \left( \Sigma_{65} \right)^2 +  
    \frac{1}{4} \lambda_{56} f^4 \left( \Sigma_{56} \right)^2.     
 \label{eq:V_quart}
\end{equation}
The first term in \Eq{eq:V_quart} breaks $SO(6)_A \times SO(6)_B$ down to $SO(5)_{A6}\times SO(5)_{B5}$, where $SO(5)_{An}$ are transformations that do not act on the $n$th row or column.  This symmetry allows $\sigma$ to get a potential but all other fields are protected.  Specifically, the $SO(5)_{A6}$ protects $h_1$ while $SO(5)_{B5}$ protects $h_2$. 
 Similarly, the second term of  \Eq{eq:V_quart} breaks the global symmetry down to $SO(5)_{A5}\times SO(5)_{B6}$.  Therefore, one loop quadratic divergences, which can only be proportional to one coupling, do not give a potential to the Higgses.  
  
In combination, the two terms of \Eq{eq:V_quart} break the global symmetry down to $SO(4)_A \times SO(4)_B$.  This means that the gauge and custodial symmetries are not explicitly broken, which is an important consistency check.  Furthermore, the fields in the upper $4\times4$ block of $\Pi$  do not get a potential from these interactions.  Plugging the parameterization of \Eq{eq:sig} into \Eq{eq:V_quart} and expanding the exponentials, we get a familiar little Higgs quartic structure:
\begin{equation} 
V_{\rm quartic} =  \frac{\lambda_{65}}{2} \left( f \, \sigma - 
                        \frac{1}{\sqrt{2}} h_1^T h_2 +  \ldots \right)^2 +
                        \frac{\lambda_{56}}{2} \left( f \, \sigma + 
                        \frac{1}{\sqrt{2}} h_1^T h_2 +  \ldots \right)^2.
\label{eq:quart_field}
\end{equation}
This potential generates a mass for $\sigma$,
\begin{equation}
m_\sigma^2 = (\lambda_{65} + \lambda_{56}) f^2 \ ,
\label{eq:sigma_mass}
\end{equation}
but no mass for the Higgses.  Furthermore, while each individual term appears to generate a quartic for the Higgses, it can be eliminated by a field redefinition 
$\sigma \rightarrow \sigma \pm \frac{h_1^T h_2}{\sqrt{2} f}$.   
In the presence of both terms, integrating out $\sigma$ at tree level yields
\begin{equation}
V_{\rm quartic} = \frac{ \lambda_{56} \lambda_{65}}{ \lambda_{65} + \lambda_{56} } 
                        \left( h_1^T h_2 \right)^2 = \frac{1}{2}\lambda_0  \left( h_1^T h_2 \right)^2 .
\label{eq:quartic}
\end{equation}
This is in the desired form of a collective quartic potential, as it is proportional to two different couplings.  Note that the potential depends only on the combination $h_1^T h_2$ which vanishes when either of the Higgs vevs is zero. A sufficiently heavy Higgs then requires  $\tan \beta \equiv \langle h_1 \rangle / \langle h_2 \rangle \sim 1 $. This will be discussed more in \Sec{sec:scal_pot}.

One might worry that the singlet $\sigma$ is ``dangerous" \cite{Schmaltz:2008vd} in that it could develop a divergent tadpole from radiative corrections. In accordance with the nonlinear symmetries, such a tadpole would be accompanied by a disastrously large mass for the Higgses $f^2(f \, \sigma \pm h_1^T h_2/\sqrt{2})$. To see that no tadpole is generated note that the Higgs potential in \Eq{eq:V_quart} respects an exact discrete symmetry
$\Sigma \rightarrow K \Sigma K$ where $K = \mathrm{diag}(1,1,1,1,1,-1)$ under which $\sigma \rightarrow - \sigma$ and $h_2 \rightarrow -h_2$.  This symmetry protects the singlet from becoming dangerous while allowing the collective quartic in \Eq{eq:quart_field}.  This discrete symmetry will be softly broken in \Eq{eq:Bmu} below.

In \App{sec:CWhiggs}, we explicitly verify that one loop diagrams involving the couplings in \Eq{eq:V_quart} do not  generate quadratically divergent masses or a tadpole for $\sigma$.  They do, however, generate logarithmically divergent masses
\begin{equation}
\frac{\lambda_{65} \lambda_{56} f^2}{16 \pi^2} \log \left( \frac{\Lambda^2}{\mu^2} \right)
     \left( h_1^T h_1 + h_2^T h_2  \right)  =
\frac{\lambda_0 m_\sigma^2}{32 \pi^2} \log \left( \frac{\Lambda^2}{m_\sigma^2} \right)
      \left( h_1^T h_1 + h_2^T h_2  \right), 
\label{eq:one-loop}
\end{equation}
where the renormalization scale $\mu$ should be chosen to minimize the finite corrections to the potential. This is accomplished for $\mu=m_\sigma$, the mass of the $\sigma$ given in \Eq{eq:sigma_mass}. Radiative corrections also generate additional quartic couplings as discussed in \App{sec:CWhiggs}, but these are numerically small and unimportant.  

\subsection{Breaking Additional Global Symmetries}

In addition to generating a collective Higgs quartic, we must also lift any remaining flat directions.  In particular, in the absence of gauge interactions, the scalars in the upper $4\times4$ block of $\Pi$ would be exact Goldstone bosons.   The gauge symmetries will explicitly break the global symmetry and give mass to some of the bosons, but not all of them, so we add the following small symmetry breaking term:
\begin{equation}
- \frac{f^2}{4} \, {\rm tr} \left[
  \left( \begin{array}{ccc}
    m_{4}^2 \, \one_4 & 0 & 0 \\
    0 & m_5^2 & 0 \\
    0 & 0 & m_6^2
  \end{array} \right) 
\Sigma \right],
\label{eq:soft_mass}
\end{equation}
where $\one_4$ is the $4\times4$ unit matrix.  This operator explicitly breaks all the axial symmetries giving a positive mass to all the scalars.  

In order to destabilize the origin in field space and trigger electroweak symmetry breaking, we also add the terms
\begin{equation}
m_{56}^2 f^2 \, \Sigma_{56} +m_{65}^2 f^2 \,\Sigma_{65} \ ,
 \label{eq:Bmu}
\end{equation}
which generate a $B_\mu$--like Higgs mass, $h_1^T h_2$.  Note that the above contributions to the Higgs potential are ``soft'', as their radiative corrections only generate suppressed contributions to the Higgs potential.

\subsection{Scalar Potential}
\label{sec:scal_pot}

We can now analyze the full scalar potential in the limit that $f$ is much larger than the electroweak scale $v_{\rm EW} \simeq 246$ GeV.  We begin with $\sigma$, the heaviest field:
\begin{equation}
V_{\sigma} = \sqrt{2} f (m_{65}^2 - m_{56}^2 ) \, \sigma - 
                        (m_{65}^2 + m_{56}^2 ) h_1^T h_2 +
                        \frac{\lambda_{65} + \lambda_{56}}{2} f^2 \sigma^2 +
                        \frac{\lambda_{56} - \lambda_{65}}{\sqrt{2}} f \, h_1^T h_2 \, \sigma \ .
\end{equation}
Here, we have ignored terms more than quadratic order in $\sigma$ as well as small radiative corrections.   Integrating out $\sigma$ at tree level, we obtain the Higgs quartic in \Eq{eq:quartic} and the operator
\begin{equation}
 -\frac{\lambda_{65} m^2_{56} + \lambda_{56} m^2_{65} }{\lambda_{65} + \lambda_{56}} \, 2 \,h_1^T h_2 \ .
\end{equation}

To analyze the scalar potential below $f$, we use the fact that the operator in \Eq{eq:soft_mass} gives a positive mass to $\eta$ and $\phi$ so there is a stable minimum at the origin of those fields.  Therefore, we need only find the minimum for the Higgs doublets.   We take the following phenomenological potential
\be
V_{\rm higgs} = \frac{1}{2} m_1^2 \, h_1^T h_1 + \frac{1}{2} \, m_2^2 h_2^T h_2 - B_\mu \, h_1^T h_2 
        + \frac{\lambda_0}{2} (h_1^T h_2)^2,
\label{eq:higgs_pot}
\ee
where we have dropped additional small Higgs quartic operators that are only generated by loop effects.  It is straightforward to determine the parameters in this potential in terms of the fundamental parameters of the theory at tree level. This parametrization is more convenient since the connection to the physical Higgs masses is more direct.  Radiative corrections such as those discussed in \Sec{sec:collectivequartic} contribute to the coefficients of the potential terms, and we will discuss additional large one loop contributions in \Secs{sec:modulargauge}{sec:top_yukawa}.

Using $SU(2)$ gauge invariance, we can choose that only the first real component of $h_1$ gets a vev.  The potential minimization equations then give a stable minimum where the only component of $h_2$ that gets a vev is the first one.  This means that the minimum will not break electromagnetism or $CP$.  Since the quartic has a flat direction if either of the $h_i=0$, both $m_1^2$ and $m_2^2$ must be positive to keep the potential bounded from below.  For the origin to be unstable and electroweak symmetry to be broken, $B_\mu > m_1 m_2$ is required.  The sign of $B_\mu$ can always be made positive by a field redefinition. We then find that the vevs of the first components are given by
\begin{eqnarray}
\langle h_{11} \rangle^2 = \frac{1}{ \lambda_0} \frac{m_2}{m_1} \,
                 (B_\mu - m_1 m_2 ) \ ,    \nonumber\\
\langle h_{21} \rangle^2 = \frac{1}{ \lambda_0} \frac{m_1}{m_2} \,
                 (B_\mu - m_1 m_2 ) \ ,
\label{eq:VEV}
\end{eqnarray}
where $h_{ij}$ is the $j$th real component of the $i$th Higgs doublet.  The generator which leaves the vacuum invariant and corresponds to electromagnetism is proportional to $T^3_L + T^3_R$.  Including small radiative quartic terms and $v_{\rm EW}/f$ corrections to the potential will induce small corrections to these vevs, but the general structure will be preserved.  

These vevs can be expressed in terms of the parameters $v_{\rm EW}$ and $\tan \beta$:
\begin{equation}
v_{\rm EW}^2 \equiv \langle h_{11} \rangle^2 +\langle h_{21} \rangle^2  = 
\frac{1}{ \lambda_0} \left( \frac{m_1^2 + m_2^2}{m_1 m_2} \right)  (B_\mu - m_1 m_2 )
\simeq ( 246 \, {\rm GeV})^2,
\label{eq:electroweak_vev}
\end{equation}
\begin{equation}
\tan \beta \equiv \frac{\langle h_{11} \rangle}{\langle h_{21} \rangle} = \frac{m_2}{m_1} \ .
\label{eq:tanbeta}
\end{equation}
From the radiative corrections due to top loops discussed in \Sec{sec:top_yukawa}, we expect $m_2 > m_1$ so that $\tan \beta > 1$. With our quartic potential, the mass-squared of the light Higgs boson scales as $M_{h^0} \sim 1/\tan \beta$ for large $\tan \beta$. Thus to clear the lower bound on the Higgs mass from LEP we require $\tan\beta \sim \mathcal{O}(1)$.  The tree level masses of the physical Higgs modes are discussed in \App{sec:physhiggs}.  By custodial symmetry, the pseudoscalar $A^0$ and charged Higgs $H^\pm$ are degenerate at tree-level
\begin{equation}
M_{A^0}^2 =M_{H^\pm}^2= m_1^2 + m_2^2 \ .
\label{eq:Amass_body}
\end{equation}
In much of the parameter space (the ``decoupling limit" \cite{Gunion:2002zf}), these states are heavier than the SM-like Higgs $h^0$ 
\begin{equation}
M_{h^0}^2 \simeq  \lambda_0 v_{\rm EW}^2 \sin^2 2\beta \ ,
\label{eq:hmass_body}
\end{equation}
and somewhat lighter than the heavy Higgs $H^0$
 \begin{equation}
M_{H^0}^2 \simeq M_{A^0}^2 + \lambda_0 v_{\rm EW}^2 \cos^2 2\beta \ .
\label{eq:Hmass_body}
\end{equation}

\subsection{Fine Tuning}
\label{sec:fine-tuning}

We are now in position to calculate whether scalar self-interactions contribute to fine-tuning in our model.  Generically, a measure of fine-tuning is the stability of the electroweak symmetry breaking Higgs vev $v^2_{\rm EW}$ under radiative corrections \cite{Barbieri:1987fn,Casas:2005ev}
\be
\Psi = \left| \frac{\delta v_{\rm EW}^2}{v_{\rm EW}^2} 
 \right| \ .
\label{eq:General_fine_tuning}
\ee
Here $\delta v_{\rm EW}^2$ denotes the shift in $v^2_{\rm EW}$ due to quantum corrections. 
The full fine-tuning of the model includes a sum over contributions from all such quantum corrections. In practice, fine-tuning is dominated by a few one loop diagrams, and we will define fine-tuning as the maximum contribution to \Eq{eq:General_fine_tuning}. In our model, the maximum contributions in each sector will come from radiative corrections to the Lagrangian parameter $m_1^2$.\footnote{This is because the gauge and quartic coupings treat $m_1^2$ and $m_2^2$ symmetrically, but the top Yukawa only affects $m_1^2$.}  Thus 
\be
\Psi \simeq \left| \frac{\partial v_{\rm EW}^2}{\partial m_1^2} \frac{\delta m_1^2}{v_{\rm EW}^2}  \right|
\simeq 2 \sin^2 \beta\, \frac{\delta m_1^2}{M_{h^0}^2}\ , 
\label{eq:fine_tuning}
\ee
where the last equality assumes the decoupling limit (see~\Sec{sec:scal_pot}) in which the SM Higgs is light compared with the other Higgs states. 
We will be interested in a range of Higgs masses. For lighter Higgs masses near 115 GeV we will find approximately 10\% fine-tuning. For intermediate masses $M_{h^0} \simeq$ 250 GeV, fine-tuning disappears and the approximation in \Eq{eq:fine_tuning} is no longer entirely accurate, though we will continue to use it because of its simplicity.  Such Higgs masses are actually preferred in our model because we find better fits to the precision electroweak data than in the SM with $M_{h^0}=$ 115 GeV.

For numerical estimates of fine-tuning here and throughout we take $f= 1$ TeV and  $\Lambda\simeq 4\pi f$.  We also use $M_{h^0} \simeq 250$ GeV and $\tan \beta \simeq \sqrt{3}$.   In the decoupling approximation, we then find $\lambda_0 \simeq 1.4$ using \Eq{eq:hmass_body} which implies $m_\sigma \simeq 1.7$ TeV.  Calculating $\Psi$ using \Eq{eq:one-loop}, we obtain $\Psi \simeq 2$, so there is $\sim$~50\% tuning in the quartic sector, which is essentially no fine-tuning.

\section{Gauge Sector}
\label{sec:gauge}

In the previous section, we only gauged the diagonal generators in $SO(4)_A \times SO(4)_B$. $W$ and $Z$ loops then contribute one loop quadratically divergent mass terms to the Higgs potential
\begin{equation}
\frac{9 \, g_{\rm EW}^2  \Lambda^2}{128 \pi^2} 
\left( h_1^T h_1 +  h_2^T h_2 \right).
\label{eq:gauge_quad}
\end{equation}
These mass terms are of order $f$ and fine-tuning of the $B_\mu$ parameter against the $m_i^2$ in \Eq{eq:electroweak_vev} would be required in order to maintain the hierarchy $v_{\rm EW} \ll f$.

\begin{figure}
\center{\includegraphics[scale=0.8]{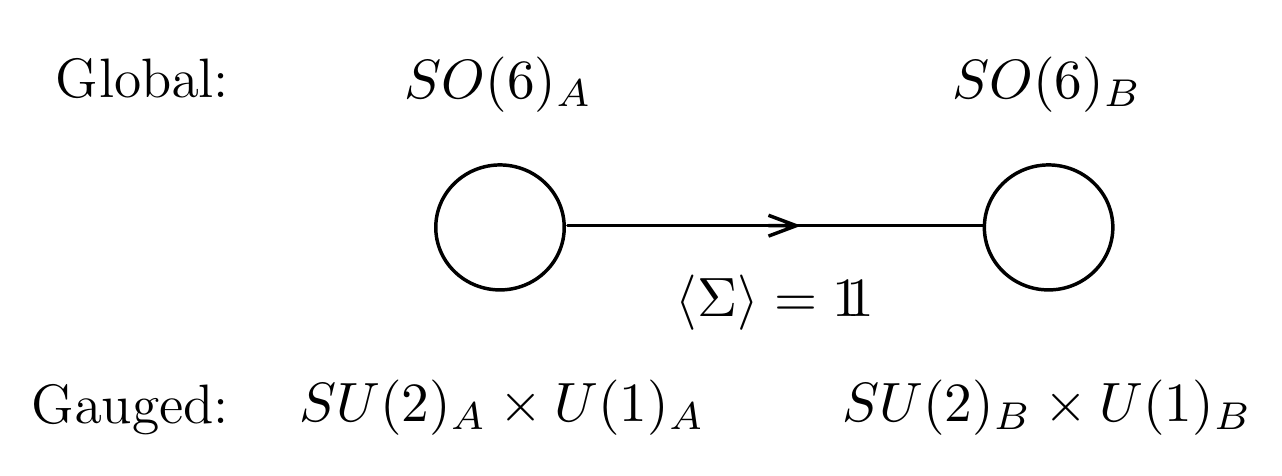}}
\caption{The traditional way to cut off gauge divergences for the model in \Fig{fig:moosediagram}.  We simply gauge the SM gauge group separately at the two sites.}
\label{fig:tradmoosediagram}
\end{figure}

In order to eliminate this quadratic divergence and reduce fine-tuning, we want to introduce the gauge couplings using collective symmetry breaking.  If we were building a traditional little Higgs model, this would be straightforward, and we explain how to implement this in the next paragraph. However, as was shown in \Refs{Csaki:2002qg, Csaki:2003si} and others, precision electroweak constraints force the masses of gauge boson partners to be above several TeV, implying that $f$ would be too large to mitigate fine-tuning in the fermion sector.  Therefore, we introduce a new method in \Sec{sec:modulargauge} to decouple the masses of gauge partners from $f$.

To implement the traditional gauge structure shown in \Fig{fig:tradmoosediagram},  we would separately gauge the generators in $SO(4)_A$ and $SO(4)_B$.   Gauging a subgroup of $SO(6)_A$ breaks some of the shift symmetries protecting the Higgs potential, but the PNGBs bosons are still protected from getting a potential by $SO(6)_B$.  The same argument applies to quantum corrections from the breaking of $SO(6)_B$. Thus any radiative corrections which contribute to the Higgs potential must be proportional to both gauge couplings.  One loop quadratically divergent diagrams are proportional to either $g_A^2$ or $g_B^2$ but not both, thus no potential is generated from gauge interactions at the one loop quadratically divergent level.   In this traditional little Higgs model, the relevant $\phi$ and $\eta$ Goldstone bosons would be eaten, giving masses to heavy $W'$ and $Z'$ bosons of order $g_{\rm EW} f$.  These heavy gauge bosons would effectively regulate the quadratic divergence from \Eq{eq:gauge_quad}.  

\subsection{A Modular Collective Gauge Coupling}
\label{sec:modulargauge}

\begin{figure}
\center{\includegraphics[scale=0.8]{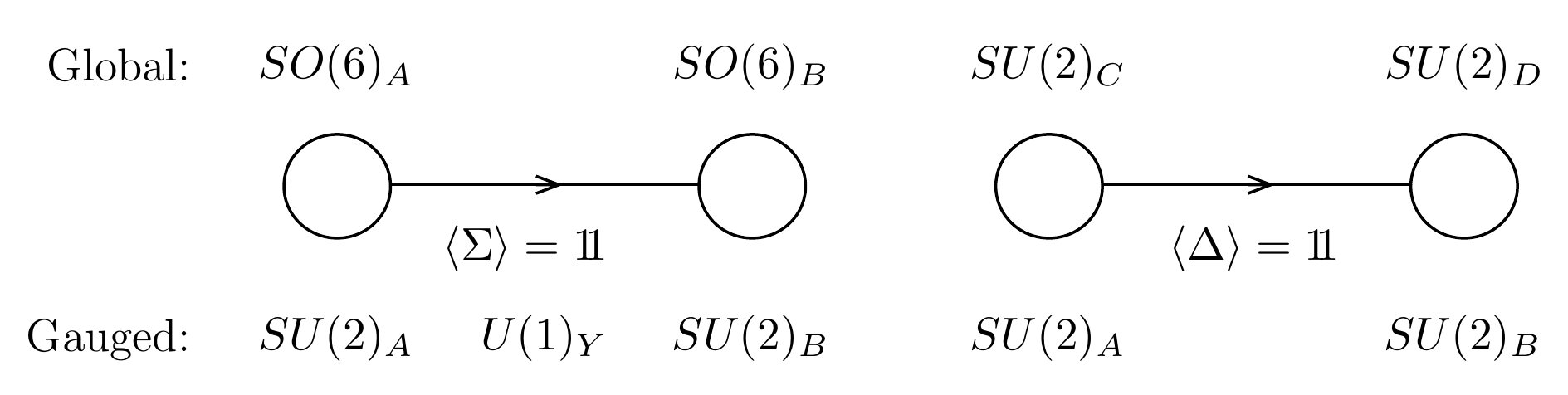}}
\caption{A modular way to cut off the gauge divergences in the model of \Fig{fig:moosediagram}. We introduce two new global symmetries and a new field $\Delta$ which transforms under them.  $\Delta$ and $\Sigma$ transform under the same $SU(2)$ gauge symmetries.  The decay constant of $\Delta$, $F$, is greater than that of $\Sigma$, so at scales below $F$, this moose reduces to \Fig{fig:moosediagram}.  We do not employ a collective breaking of hypercharge and simply gauge the diagonal combination in $SO(6)_A\times SO(6)_B$.}
\label{fig:newmoosediagram}
\end{figure}

Instead of taking the traditional approach for building a collective gauge coupling, we will take a modular one which decouples the dynamics which cuts off gauge boson divergences from the dynamics that generates the quartic and fermion interactions.\footnote{Separation of the gauge boson and top partner masses with different $f$'s also arises in a limit of the simple group models \cite{Kaplan:2003uc, Schmaltz:2004de, Skiba:2003yf}. However in these models, the top partners remain heavier than gauge boson partners for reasonable hierarchies of scales $F > f$.  See also \cite{Barcelo:2007if,Barcelo:2008je}.}   This still requires separately gauging the generators in $SO(4)_A$ and $SO(4)_B$, but the $\phi$ and $\eta$ PNGBs will remain uneaten, changing the collider phenomenology.

The setup we envision is shown in \Fig{fig:newmoosediagram}. We start with two independent sigma models: $\Sigma$ breaks a global $SO(6)_A \times SO(6)_B$ symmetry down to the diagonal at the scale $f$ as before, and the new field $\Delta$ breaks a new global $SU(2)_{C} \times SU(2)_{D}$ symmetry to the diagonal at a higher scale $F>f$. To connect these two independent sigma models, we now gauge the $SU(2)_{LA}$ symmetry in $SO(6)_A$ and $SU(2)_C$ with the same gauge group $SU(2)_A$. Similarly, we gauge $SU(2)_{LB}$ and $SU(2)_D$ with the same gauge bosons $SU(2)_B$. Note that the two sigma models break two separate global $SU(2)\times SU(2)$ symmetries to the diagonal $SU(2)$, thus we expect two sets of light triplet NGBs. However, the two $SU(2)\times SU(2)$ global symmetries are gauged with the same $SU(2)_A \times SU(2)_B$ gauge bosons, thus only one set of NGB triplets is eaten and the other remains light.

The vevs of $\Sigma$ and $\Delta$ both contribute to the masses of the off-diagonal $SU(2)$ gauge bosons which are now proportional to $\sqrt{f^2+F^2}\sim F$. The gauge bosons partners can be decoupled by taking $F$ larger than $f$. We will determine the allowed values of $F$ in \Sec{sec:pew}, but the main point is that $F$ can be made large enough to avoid precision electroweak constraints without causing any fine-tuning in the quartic or top sectors.

More concretely, the gauge invariant non-linear sigma kinetic terms are
\begin{equation}
{\cal L} = \frac{f^2}{8} {\rm tr}\left(D_\mu \Sigma^\dagger D^\mu \Sigma \right) +
                 \frac{F^2}{4} {\rm tr}\left(D_\mu \Delta^\dagger D^\mu \Delta \right) , 
\label{eq:kinetic}
\end{equation}
where the coefficients are determined to ensure that the NG modes have canonical kinetic terms, and 
\begin{equation}
D \Sigma = \partial \Sigma + i g_A A_{1}^{(6)} \Sigma - i g_B \Sigma \, A_{2}^{(6)}, \qquad
D \Delta = \partial \Delta + i g_A A_{1}^{(2)} \Delta - i g_B \Delta \, A_{2}^{(2)},
\end{equation}
\begin{equation}
A_i^{(6)} = A_i^a T^a, \qquad A_i^{(2)} = A_i^a \frac{\tau^a}{2},
\end{equation}
where $T^a$ are given in \App{sec:grouptheory} and $\tau^a$ are the Pauli matrices, $g_A$ and $A_1$ are the gauge coupling and field associated with $SU(2)_{LA}$, and $g_B$ and $A_2$ are associated with $SU(2)_{LB}$.  We parameterize the NG modes in $\Delta$ as
\begin{equation}
\Delta = F e^{2i \Pi_d / F}, \qquad \Pi_d = \chi_a \frac{\tau^a}{2} \ .
\end{equation}
When $\Sigma$ and $\Delta$ get vevs, one gauge boson gets heavy while the other remains massless,
\begin{equation}
A_0^a = \cos \theta_g A_1^a + \sin \theta_g A_2^a \ ,  \qquad
A_H^a =\sin \theta_g A_1^a - \cos \theta_g A_2^a \ ,
\label{eq:gauge_eigen}
\end{equation}
where 
\be
\tan \theta_g  = \frac{g_A}{g_B} \ .
\label{eq:theta-g}
\ee
The mass of the heavy gauge boson is given by\footnote{\Eqs{eq:gauge_eigen}{eq:heavy_gauge} will have corrections of order $v^2/F^2$ when the Higgses get vevs.  }
\begin{equation}
m^2_{W'} = \frac{1}{4}(g_A^2 + g_B^2)(f^2 + F^2) \ .
\label{eq:heavy_gauge}
\end{equation}
The unbroken $SU(2)_{\rm EW}$ electroweak gauge coupling is
\begin{equation}
\frac{1}{g_{\rm EW}^2} = \frac{1}{g_A^2} + \frac{1}{g_B^2} \ .
\label{eq:gauge_coupling}
\end{equation}
The heavy gauge boson eats one linear combinations of the triplets, $\phi$ and $\chi$, while leaving one remaining as the physical triplet:
\begin{equation}
\phi_{\rm eaten}^a = \frac{1}{\sqrt{ f^2 + F^2}}(f \, \phi^a + F \chi^a) \ ,  \qquad
\phi_{\rm physical}^a = \frac{1}{\sqrt{ f^2 + F^2}} ( F \phi^a -  f \,  \chi^a ) \ .
\end{equation}
We will work in unitary gauge where the eaten triplet does not appear.  

In \App{sec:CWgauge}, we show that radiative corrections from the gauge sector do not destabilize the Higgs potential in \Eq{eq:higgs_pot}.  The leading correction to the Higgs boson and triplet masses is
\begin{eqnarray}
\frac{9 \, g_A^2 \, g_B^2}{512 \pi^2} \log \left( \frac{\Lambda^2}{m_{W'}^2} \right)  
 \left(  f^2 + F^2 \right) \left( h_1^T h_1 + h_2^T h_2 
 +  \frac{8}{3} \phi_a \phi^a \right) \nonumber\\
  = \frac{9 \, g_{\rm EW}^2 m_{W'}^2}{128\pi^2}  \log \left( \frac{\Lambda^2}{m_{W'}^2} \right)
  \left( h_1^T h_1 + h_2^T h_2 
 +  \frac{8}{3} \phi_a \phi^a \right)  .
 \label{eq:gauge_mass}   
\end{eqnarray}
From this we see that the quadratic divergence generated by the light gauge bosons is cut off by the heavy ones.  Namely, interpreting the cutoff in \Eq{eq:gauge_quad} as the mass of the heavy gauge bosons given in \Eq{eq:heavy_gauge}, we reproduce \Eq{eq:gauge_mass} up to logs of the new cutoff.  We should now discuss how much fine-tuning the gauge sector generates in the Higgs potential, but we will postpone this discussion to \Eq{eq:gauge_finetuning} of \Sec{sec:pew} after we have done a detailed analysis of the constraints on the mass of the heavy gauge bosons.  The results of \Sec{sec:pew} will be that the heavy gauge bosons must have a mass between 1.5 and 3 TeV or greater depending on details of the fermion charge assignments, and that the gauge sector will also not generate any fine-tuning in this model.  

The careful reader will have noted that the operator in \Eq{eq:soft_mass} is not gauge invariant once we gauge  $SU(2)_A$ and $SU(2)_B$ in $SO(4)_A \times SO(4)_B$ separately.  The operator must now involve both $\Delta$ and $\Sigma$ to reproduce the mass terms in \Eq{eq:soft_mass}. 
\begin{equation}
- \frac{f^2}{4} m^2_{4} \, {\rm tr}\left( \Delta^\dagger M_{26} \Sigma M_{26}^\dagger +
             \Delta M_{26} \Sigma^\dagger M_{26}^\dagger \right) - 
             \frac{f^2}{4} \left( m_5^2 \Sigma_{55} + m_6^2 \Sigma_{66} \right)      \ .  
\label{eq:univ_mass}
\end{equation}
Here, $M_{26}$ is a $2\times 6$ matrix which preserves all the gauge symmetries\footnote{We have gauged the diagonal $T^3_R$ generator as hypercharge, see \Sec{sec:hypercharge}.} and is given in \App{sec:grouptheory}.  These operators give the same masses for the fields as \Eq{eq:soft_mass}, up to corrections of order $f^2/F^2$.

It is interesting to note that this way of implementing collective symmetry breaking in the gauge sector works with almost any little Higgs model.  To any model with an $SU(2)_A\times SU(2)_B$ gauge structure one simply adds an additional sigma model $\Delta$ which breaks $SU(2)_A\times SU(2)_B$ at the higher scale $F$, thus raising the gauge boson partner masses and alleviating precision electroweak constraints without altering much else. The tell-tale phenomenological signature of this approach is the appearance of an uneaten light $SU(2)$ triplet of PNGBs.

Philosophically, this construction is similar to the vector limit of QCD~\cite{Piai:2004yb} and the intermediate Higgs~\cite{Katz:2005au} in the sense that dynamics above the scale $f$ can be used to regulate the gauge divergences.  The difference is that here, nothing about the $\Sigma$ non-linear sigma model had to be modified (except the gauge interactions).  In particular, the operators that generated the quartic are identical with or without the $\Delta$ field, and we will see that the fermion structure is also identical.   Because the $\Sigma$ non-linear sigma model requires a UV completion at $\Lambda \sim 4 \pi f$, this setup strictly speaking only valid if $F < \Lambda$.  One might be concerned about how to explain the ``little hierarchy'' between the scales $f$ and $F$, but one could certainly imagine a cascading strongly-coupled sector that dynamically yields two different effective decay constants.  In any case, we are only interested in describing the low energy dynamics below $\Lambda \sim 4 \pi f$, where this modular collective gauge mechanism is simple and robust.  

\subsection{Custodial Symmetry}

Independent of the collective gauge coupling mechanism, it is important to check that the low energy dynamics has an approximate custodial $SU(2)$ symmetry.  By construction, the Higgses transform as $\mathbf{4}$s of $SO(4)$, so before we introduce custodial-violating hypercharge or Yukawa couplings, the Higgs potential respects custodial $SU(2)$.  Crucially, the actual vacuum also respects custodial $SU(2)$, even accounting for $v_{\rm EW} / f$ corrections to the Higgs potential.  

In order to check that the vacuum preserves custodial symmetry, we can replace all the fields with their vevs (only $h_{i1}$ get vevs) in \Eq{eq:sig} and then plug that into the kinetic term in \Eq{eq:kinetic} to determine the full mass matrix of the gauge bosons.  The mass terms for the gauge bosons are given by 
\be
\bigg( A_1^a \; A_2^a \bigg)
 \frac{1}{4} \left( \begin{array}{cc}
   g_A^2(f^2+F^2) & 
   -g_A g_B \left(F^2+f^2 \cos^2 \left(\frac{ v_{\rm EW}}{\sqrt{2} f}\right)\right)\\
   -g_A g_B \left(F^2+f^2 \cos^2 \left(\frac{v_{\rm EW}}{\sqrt{2} f}\right)\right) 
   &  g_B^2 (f^2+F^2)
  \end{array} \right)
  \left( \begin{array}{c}
   A_1^a \\ A_2^a \end{array}\right).
\ee
The important thing to note about this is that it does not depend on the $SU(2)$ index $a$, so before we add hypercharge, the three light gauge bosons are degenerate.  This shows that at all orders in the $v_{\rm EW} / f$ expansion, custodial symmetry is preserved and there are no new tree-level contributions to the $T$ parameter in our model, as long as the vacuum described in \Sec{sec:scal_pot} isn't destabilized.  Expanding the eigenvalues of the mass matrix to leading order in $v_{\rm EW}/f$, we recover \Eq{eq:heavy_gauge} for the larger eigenvalue, and the mass of the light gauge bosons is given by  
\begin{equation}
m^2_W = \frac{g_{\rm EW}^2}{4} v_{\rm EW}^2 + \mathcal{O}(v^4/f^2) \ .
\label{eq:WZ_mass}
\end{equation}
This means that in our normalization $v_{\rm EW}= \sqrt{v_1^2 + v_2^2} \simeq 246$ GeV.

\subsection{Hypercharge}
\label{sec:hypercharge}

In \Sec{sec:modulargauge}, we softened quadratic divergences from $SU(2)_L$ gauge interactions by introducing a new sigma field $\Delta$.  In principle, one could also do the same for the $U(1)_Y$ gauge coupling by introducing a third sigma field $\Delta'$.  This would require having two hypercharge couplings which both break the custodial symmetry.  A more custodially symmetric way to accomplish this would be to enhance the global symmetry which $\Delta$ transforms under to $SO(4)\times SO(4)$ and then gauge one full $SO(4)$ while only gauging $SU(2)\times U(1)$ at the other site.   This means that there is only one spurion which breaks the custodial symmetry, and by making the $SO(4)$ gauge coupling somewhat large, this spurion can be made roughly the same size as the SM hypercharge coupling.  

In this paper, we choose a different route.  Given the relative smallness of the hypercharge coupling and the fact that we are only working in the theory below $\Lambda \sim 4 \pi f$, gauge bosons which cut off the hypercharge quadratic divergence worsen precision electroweak physics without much gain in fine-tuning.  Therefore, we simply gauge hypercharge as the diagonal combination of $T_R^3$ in the $\Sigma$ sector, while leaving the $\Delta$ sector unchanged.  The normalization is such that $h_1$ and $h_2$ have hypercharge 1/2 in the notation of \Eq{eq:higgsdoublet} in \App{sec:grouptheory}, while the singlet formed by $\eta_1$ and $\eta_2$ has hypercharge 1.  All other fields are neutral.  Since no collective symmetry breaking is employed, the quadratically divergent mass terms generated by hypercharge are 
\begin{equation}
\frac{3 g_Y^2 \Lambda^2}{32 \pi^2}
\left[ \eta_1^2 + \eta_2^2 + \frac{1}{4}( h_1^T h_1 + h_2^T h_2 ) \right].
\label{eq:hyp_mass}
\end{equation}
Because of this, we expect to the complex singlet formed by $\eta_1$ and $\eta_2$ to be substantially heavier than $\eta_3$ which only gets a mass from the operator in \Eq{eq:univ_mass}.  Calculating the fine-tuning parameter from \Eq{eq:fine_tuning}, we get
\be
\Psi_{\rm Y} = \frac{3\, g_Y^2 \sin^2 \beta \,\Lambda^2}{32 \,\pi^2 M_{h^0}^2 } \ .
\ee
Taking $g_Y\simeq 0.3$ and again taking $\tan \beta \simeq \sqrt{3}$ and $M_{h^0} \simeq 250$ GeV, we find that $\Psi_{\rm Y} \simeq 2$, so the the hypercharge sector does not contribute to fine-tuning even with the quadratic divergence.

There are additional quadratically divergent contributions to the Higgs potential which we calculate in \App{sec:CWhyp}.  These interaction terms do not destabilize the Higgs potential analyzed in \Sec{sec:scal_pot}, and the small modifications  do not change the vevs by very much.  While some of the Higgs quartic interactions in the radiative potential are not custodially symmetric, they are proportional to hypercharge and similar to the terms which would be present in any two Higgs doublet model, so they are not in conflict with precision electroweak constraints.  

\section{Fermion Sector}
\label{sec:fermion}

Having successfully implemented collective quartic and gauge couplings, we now implement the fermion sector.  We begin by understanding the $SO(6)$ representations that fermions will live in.  Consider a fundamental (i.e.\ $\mathbf{6}$) of $SO(6)$; it contains two singlets and two $SU(2)_L$ doublets.   To build Yukawa interactions, we will need fermions that transform either under $SO(6)_A$ or $SO(6)_B$.  The $SO(6)_A$ fundamental $Q$ decomposes as 
\begin{equation}
Q^T =
    \left( \begin{array}{cccccc}
    \frac{1}{\sqrt{2}}( -Q_{a1} - Q_{b2})  &
    \frac{i}{\sqrt{2}}( Q_{a1} - Q_{b2}) &
    \frac{1}{\sqrt{2}}( Q_{a2} - Q_{b1}) &
    \frac{i}{\sqrt{2}}( Q_{a2} + Q_{b1}) &
    Q_5 &
    Q_6
  \end{array} \right)\ ,
  \label{eq:fermionA}
\end{equation}
where $Q_a=(Q_{a1},Q_{a2})$ forms an $SU(2)_L$ doublet with hypercharge $-\frac12$, while $Q_b$ is a doublet with hypercharge $\frac12$. Together $Q_a$ and $Q_b$ form a bi-doublet under $SU(2)_L\times SU(2)_R \equiv SO(4)$, and couplings which respect custodial symmetry must treat them symmetrically. The $Q_5$ and $Q_6$ fields are singlets under $SO(4)$.  

We wish to identify $Q_a$ with an SM quark doublet with hypercharge $\frac16$.\footnote{In principle, we could have used a $\mathbf{4}$ instead of a $\mathbf{6}$ for the fermion representations, but the $\mathbf{4}$ encounters difficulties with custodial $SU(2)$.} Thus, we must modify the definition of hypercharge acting on the fermions, and we do so by gauging a linear combination of $T^3_R$ and a global symmetry of the fermion sector $U(1)_X$. The hypercharge generator is then
\be
T_Y = T^3_R + T_X \ .
\ee
Choosing the $U(1)_X$ charge of $Q$ to be $\frac23$ reproduces the correct hypercharge for $Q_a$. 
The $U(1)_X$ charges of all fermion representations are given below in \Tabs{table:charge_top}{table:charge_light}.

For the fundamental of $SO(6)_B$, we use a slightly different notation in which the two $SU(2)_L$ doublets are switched:  
\begin{equation}
(U^c)^T =
    \left( \begin{array}{cccccc}
    \frac{1}{\sqrt{2}}( -U^c_{b1} - U^c_{a2})  &
    \frac{i}{\sqrt{2}}( U^c_{b1} - U^c_{a2}) &
    \frac{1}{\sqrt{2}}( U^c_{b2} - U^c_{a1}) &
    \frac{i}{\sqrt{2}}( U^c_{b2} + U^c_{a1}) &
    U^c_5 &
    U^c_6
  \end{array} \right) \ .
  \label{eq:fermionB}
\end{equation}
The SM up-type singlet will live in the fifth component of $U^c$.  This switched notation has the virtue that fields with identical indices have the correct quantum numbers to obtain Dirac masses. For example, $Q_5$ can have a mass with $U^c_5$ and $Q_a$ with $U^c_a$ (assuming $U(1)_X$ charge $-\frac23$ for $U^c$). 

\subsection{Top Yukawa Coupling}
\label{sec:top_yukawa}

The biggest coupling in the fermion sector is the top Yukawa coupling. We will introduce this coupling using collective symmetry breaking in such a way that one-loop radiative corrections to the Higgs masses from the top and top partners are finite and proportional to the top partner masses.  In order to minimize top partner masses (and therefore radiative corrections to the Higgs mass), we will adopt the ``bestest" structure for the top Yukawa coupling which minimizes the mass of the top partners for fixed top Yukawa coupling.   Similar constructions have been used previously in \Refs{Kaplan:2003uc, Schmaltz:2004de, Cheng:2006ht}, and we mention other less ideal top Yukawa structures in \App{sec:alttop}.  The Yukawa couplings for the remaining quarks do not require special care; we will briefly discuss them in \Sec{sec:light_quarks}.

\begin{table}
\begin{center}
\begin{tabular}{c|c|c|c|c}
 &  $SO(6)_A$ & $SO(6)_B$ & $SU(3)_C$ & $U(1)_X$ \\
 \hline
$Q$ & $\mathbf{6}$ & $-$ & $\mathbf{3}$ & $2/3$ \\
${Q'}_a$ & $\mathbf{2}^{(*)}$ & $-$ & $\mathbf{3}$  & $2/3$ \\
$U^c$ & $-$ & $\mathbf{6}$& $\mathbf{\overline{3}}$ & $-2/3$\\
${U'_5}^c$ & $-$ & $\mathbf{1}^{(*)}$ & $\mathbf{\overline{3}}$ & $-2/3$
\end{tabular} 
\end{center}
\caption{Fermion charge assignments for the top sector. Note that the fermions $Q'_a$ and ${U'_5}^c$ form incomplete representations of $SO(6)_A$ and $SO(6)_B$ respectively.   The notation $\mathbf{2}^{(*)}$ indicates that $Q'_a$ is a doublet of $SU(2)_A$, and $\mathbf{1}^{(*)}$ indicates that ${U'_5}^c$ is a singlet of $SU(2)_B$.}
\label{table:charge_top}
\end{table}

For the top Yukawa coupling, we use the fermion multiplets $Q$ and $U^c$  given above which transform as fundamentals of $SO(6)_A$ and $SO(6)_B$, as shown in \Tab{table:charge_top}.  In addition, we use an $SU(2)_A$ doublet $Q'_a$ and a singlet ${U'_5}^c$, which can be considered as incomplete multiplets of $SO(6)_A$ and $SO(6)_B$.\footnote{In particular, we are imagining that the full $SO(6)_A \times SO(6)_B$ is a good global symmetry above the scale $\Lambda$, see e.g. \cite{Thaler:2005en}.}  Note that except for the primes, we are using the same names for these incomplete multiplets as for the corresponding components of $Q$ and $U^c$. This notation indicates that the fields with identical names mix, and one linear combination will be heavy while the other will correspond to third generation quark fields.

The collective top Yukawa coupling is
\begin{equation}
\mathcal{L}_t = y_1f \, Q^T \, S\, \Sigma \, S \, U^c + y_2 f\, {Q'_a}^T \,  \Sigma \, U^c
+ y_3 f \, Q^T \, \Sigma \, {U'_5}^c\, + \ {\rm h.c.} \ ,
\label{eq:top_yukawa}
\end{equation}
where the incomplete multiplets ($Q'_a$ and ${U'_5}^c$) are contracted with $\Sigma$ like normal $SO(6)$ multiplets but with extra components set to zero. To be concrete, 
${Q'_a}^T \rightarrow \frac{1}{\sqrt{2}} (-Q'_{a1},  i Q'_{a1}, Q'_{a2}, i Q'_{a2}, 0,0)$ and 
${U'_5}^{cT} \rightarrow (0,0,0,0,{U'_5}^c,0)$. $S$ is the $SO(6)$ matrix $S={\rm diag}(1,1,1,1,-1,-1)$.\footnote{The 6th component of $S$ could be $+1$ with minimal changes to the phenomenology.}
Note that $S^2=1$, therefore the $S$ matrices can also be moved into the exponent of $\Sigma$ where their only effect is to flip the sign of the Higgs fields $h_1$ and $h_2$.  Also, the inclusion of the $S$ matrix does not break any of the gauge symmetries.

We now explore the symmetry structure of these couplings. In the first term, it looks as if the $S$ matrices explicitly break the $SO(6)_A$ and $SO(6)_B$ symmetries.  However note that the $S$'s can be absorbed into the fermion fields by field redefinitions (which will make the $S$'s pop back up in the second and third terms). In this new basis, the  first term manifestly preserves a full $SO(6)_A\times SO(6)_B$ symmetry which protects any of the NGBs from obtaining a non-derivative coupling. 

Going back to the original basis, the second term preserves the $SO(6)_B$ symmetry which is sufficient to protect the NGBs from radiative corrections.  Understanding the symmetry left unbroken by the first two terms together is subtle. Making the two field redefinitions $\Sigma U^c \rightarrow \widetilde U^c$ and $Q^T S \Sigma S \Sigma^T \rightarrow \widetilde Q^T$, the first two terms become 
$y_1f \, \widetilde Q^T \, \widetilde U^c + y_2 f\, {Q'_a}^T \, \widetilde U^c$ which manifestly does not depend on $\Sigma$. Thus no potential for any of the NGBs can result from  these two terms alone. Note however that the kinetic terms of the fermion fields are not invariant under $\Sigma$-dependent field redefinitions. Thus the NGB interactions cannot be removed completely, but we see that the NGBs only have derivative couplings.

The third term manifestly preserves $SO(6)_A$, and a similar argument shows that any two terms involving the third coupling also cannot give a Higgs potential.  However, when all three terms are included, all of the symmetries protecting the Higgs are broken.  In particular, because of the $S$ matrices, there is no fermion field redefinition that can remove all non-derivative $\Sigma$ interactions.  Thus, we have collectively broken all the symmetries protecting the Higgs so that a top Yukawa coupling (and radiative corrections to the Higgs potential) can be generated.  The radiatively generated Higgs potential must involve all three couplings and is finite at one loop. Since the full upper $SO(4)$ subgroup of $SO(6)_B$ is preserved by all three terms, radiative correction from the top sector cannot generate a potential for $\phi$ or $\eta$ at any order.  

We can read off the electroweak symmetry preserving parts of the top partner masses by replacing $\Sigma$ in  \Eq{eq:top_yukawa} by the unit matrix
\begin{eqnarray}
\mathcal{L}_t &\supset& y_1f \left(Q_b U_b^c  +Q_6 U_6^c \right)  \nonumber \\
&+&\sqrt{|y_1|^2+|y_2|^2}\, f\, \left(\frac{y_1}{\sqrt{|y_1|^2+|y_2|^2} } Q_a+ \frac{y_2}{\sqrt{|y_1|^2+|y_2|^2} } \, Q'_a \right)\, U_a^c  \nonumber \\
&+& \sqrt{|y_1|^2+|y_3|^2}\,f\, Q_5\, \left(\frac{y_1}{\sqrt{|y_1|^2+|y_3|^2} } U^c_5 + \frac{y_3}{\sqrt{|y_1|^2+|y_3|^2} } {U'_5}^c \right)\  + \ {\rm h.c.} \ .
\label{eq:top-partner-mass}
\end{eqnarray}
We have a whole slew of top partners that can loosely be regarded as a $\mathbf{6}$ of $SO(6)$.   An $SU(2)$ doublet of quarks $T_a$ with mass $m_{T_a}=\sqrt{|y_1|^2+|y_2|^2}\, f$ (corresponding to a linear combination of $Q_a$/$Q'_a$ married with $U^c_a$); a doublet of quarks $T_b$ with mass $m_{T_b}=y_1 f$  ($Q_b$ with $U^c_b$); the singlet $T_5$ with mass $m_{T_5}= \sqrt{|y_1|^2+|y_3|^2}\, f$ ($Q_5$ with a linear combination of $U^c_5$/${U'_5}^c$); and the singlet $T_6$ with mass $m_{T_6}= y_1 f$ ($Q_6$ with $U^c_6$).

The two remaining linear combinations do not obtain a mass before electroweak symmetry breaking and correspond to the third generation quark doublet $q_3$ and the ``right-handed" top quark $u^c_3$:
\begin{eqnarray}
q_3&=& \frac{y^*_2}{\sqrt{|y_1|^2+|y_2|^2} }\,  Q_a- \frac{y^*_1}{\sqrt{|y_1|^2+|y_2|^2} }\, Q'_a \ , \nonumber \\
u^c_3&=& \frac{y^*_3}{\sqrt{|y_1|^2+|y_3|^2} }\, U^c_5 - \frac{y^*_1}{\sqrt{|y_1|^2+|y_3|^2} }\, {U'_5}^c \ .
\end{eqnarray}
Their couplings to the Higgs can be found by expanding \Eq{eq:top_yukawa} to first order in the Higgs and projecting onto the light fermions
\begin{eqnarray}
-\, 3\, \frac{ y_1 y_2 y_3}{\sqrt{|y_1|^2 + |y_2|^2}\sqrt{|y_1|^2 + |y_3|^2}}\,
\left([q_3]_1 ( h_{11} - i h_{12}) - [q_3]_2 (h_{13} + i h_{14}) \right) u_3^c \ ,
\label{eq:top_higgs}
\end{eqnarray}
where $[q_3]_i$ are the components of the quark doublet and $h_{1i}$ are the components of the Higgs.  Note that $h_{11}$ is the component that gets a vev, thus $[q_3]_1$ is the left handed top while $[q_3]_2$ is the left handed bottom. The factor of 3 in \Eq{eq:top_higgs} results from adding contributions from the three terms in \Eq{eq:top_yukawa}.\footnote{See \App{sec:alttop} for more discussion of this factor of 3.} 

To get a feel for the size of the couplings and masses consider $y=y_1=y_2=y_3$ for which \Eq{eq:top_higgs} simplifies to $y_t=\frac32\, y$. For example, assuming as before that $\tan \beta = \sqrt{3}$, we must choose
$y = \frac23 y_t=\frac23\, \frac{m_t}{v_1}\simeq \frac{2\sqrt{2}}{3\sqrt{3}}\ {m_t}/{(174~{\rm GeV})} $ to get the correct top mass. For $f=1$ TeV this implies top partner masses $m_{T_a}=m_{T_5}=\sqrt{2} y \,f \simeq 770$ GeV and $m_{T_b}=m_{T_6}=y f \simeq 540$ GeV. Moving away from equal couplings $\lambda_i$ while keeping the top quark mass fixed increases the mass of at least one of the top partners.

One loop corrections to the Higgs potential from fermion loops are calculated in \App{sec:CWfermions}.  The leading correction to the mass terms in the Higgs potential is finite and given by 
\begin{equation}
-\frac{3 f^2}{8 \pi^2}9\frac{|y_1|^2 |y_2|^2  |y_3|^2 }{|y_2|^2 - |y_3|^2}
\log \left( \frac{|y_1|^2 + |y_2|^2 }{|y_1|^2 + |y_3|^2 } \right)
h_1^T h_1 
= -\frac{3 \, m_t^2}{8 \pi^2 v_1^2}  \,  \frac{m_{T_a}^2 \, m_{T_5}^2}{m_{T_a}^2-m_{T_5}^2}
\log \left( \frac{m_{T_a}^2}{m_{T_5}^2} \right)
h_1^T h_1 \ .
\label{eq:fermion_higgs}
\end{equation}
From this we see that the heavy top partners which mix with the real top quark cut off the divergence in the top sector.  The mass term for $h_1$ is negative, but as long as there are larger tree- or loop-level positive contributions, the vacuum is not destabilized.   As shown in \App{sec:CWfermions}, the one loop potential from the fermion sector respects custodial symmetry, despite the fact that the $Q'_a$ multiplet violates $SO(4)$.

Finally, we compute the fine-tuning measure $\Psi$ defined in \Eq{eq:fine_tuning}. We get \begin{eqnarray}
\Psi_t &=& \frac{3 m_t^2 }{2\, \pi^2 v_{\rm EW}^2 M_{h^0}^2}
                     \frac{m_{T_a}^2 m_{T_5}^2}{m_{T_a}^2 - m_{T_5}^2}
                      \log \left( \frac{m_{T_a}^2}{m_{T_5}^2} \right) \nonumber \\
                      &=&  \frac{3}{4 \pi^2} \left(\frac{m_t}{174~{\rm GeV}}\right)^2\, 
                      \frac{m_{T_a}^2}{ M_{h^0}^2} \simeq 0.7 \ ,
\end{eqnarray}
where in the second line we assumed equal top partner masses $m_{T_a}=m_{T_5}$ and plugged the reference values $m_{T_a}=770$ GeV and $M_{h^0}=250$ GeV. We see that there is absolutely no fine-tuning from the ``bestest" top sector at this point in parameter space.

\subsection{Light Fermions}
\label{sec:light_quarks}

The Yukawa couplings for the remaining quarks are small, and any quadratic divergences proportional to these small couplings squared do not introduce fine-tuning to the theory.   This is even true for bottom quark loops which are proportional to $\frac{3 y_b^2}{16 \pi^2} \Lambda^2 \sim 10^{-5}\Lambda^2$.  As described below, all SM fermions will obtain their masses from the same Higgs, thus there is no $\tan \beta$ enhancement of down-type Yukawa couplings.  

We introduce the remaining fermions in a straightforward way, with charge assignments given in \Tab{table:charge_light}.  For the up-type quarks, we introduce two generations of quark doublets $q$ and singlets $u^c$ and couple them directly to the components of $\Sigma$ which contain the Higgs doublet at linear order. This is most conveniently done by writing $q$ as a 6-component object like we did for $Q'_a$ in the previous subsection, $q^T \rightarrow \frac1{\sqrt{2}}(-[q]_1,i [q]_1,[q]_2, i [q]_2,0,0)$. Here $[q]_1$ and $[q]_2$ are the up- and down-type left-handed quarks, respectively.  The up-type Yukawa coupling becomes
\begin{equation}
q_i \,  \Sigma_{i5} \, u^c  \rightarrow   \left(
[q]_1 (h_{11} - i h_{12}) - [q]_2 (h_{13} + i h_{14}) \right) u^c,
\label{eq:c_yukawa}
\end{equation}
where we have suppressed all generation indices.
Note that this term preserves the upper $SO(4)_B$ subgroup of $SO(6)_B$ and therefore cannot generate a potential for $\eta$ or $\phi$.

\begin{table}
\begin{center}
\begin{tabular}{c|c|c|c|c}
 &  $SU(2)_A$ & $SU(2)_B$ & $SU(3)_C$ & $U(1)_X$ \\
 \hline
$q$ & $\mathbf{2}$ & $-$ & $\mathbf{3}$ & $2/3$ \\
$u^c$ & $-$ & $-$ & $\mathbf{\overline{3}}$ & $-2/3$\\
$d^c$ & $-$ & $-$ & $\mathbf{\overline{3}}$ & $1/3$\\
$\ell$ & $-$ & $\mathbf{2}$ & $-$ & $0$ \\
$e^c$ & $-$ & $-$  & $-$ & $1$ 
\end{tabular}
\end{center}
\caption{Fermion charge assignments for light quarks and leptons. }
\label{table:charge_light}
\end{table}

The Yukawa couplings for down-type quarks also do not require collective symmetry breaking, and we again simply couple to the components in $\Sigma$ which contain the $h_1$ Higgs. Down-type quarks have Yukawa couplings with the charge conjugate Higgs. In $SO(4)$ notation charge conjugation is accomplished by multiplying with $-2i T_R^2$, thus
\begin{equation}
q_i \,  (-2i T_R^2 \Sigma)_{i5} \, d^c \rightarrow  \left(
[q]_1 (- h_{13} + i h_{14}) - [q]_2 (h_{11} + i h_{12}) \right) d^c\ .
\label{eq:b_yukawa}
\end{equation}
For simplicity, we imagine that the couplings in the up-sector are flavor diagonal, and thus the CKM matrix comes entirely from the down-type Yukawa couplings, though other flavor structures are certainly possible.

Since the light fermions' Yukawa couplings are just the ordinary SM Yukawa couplings, no new flavor violation is introduced  to lowest order. Expanding $\Sigma$ to higher order in fields leads to couplings of the $\phi^a$ and $\eta^a$ PNGBs to the light fermions. These couplings are harmless because they are proportional to quark masses over $f$. There is some violation of flavor universality of the top quark due to its mixing with heavy top partners, however the effects are $v_{\rm EW}/f$ suppressed and not very large. The coupling of the $Z$ to bottom quarks is not modified at tree level because the bottom quark does not mix with new fermions which have different $SU(2)\times U(1)$ quantum numbers. At one loop, modifications to $Z\rightarrow b \bar b$ are comparable to the Littlest Higgs model with $T$-parity, and small enough that it does not affect precision electroweak constraints \cite{Hubisz:2005tx}. We discuss electroweak constraints in more detail in the next section.

For the charged leptons we have a choice. We can either implement their Yukawa couplings in the same way as the down-type quarks, in which case the lepton doublets would be charged under $SU(2)_A$.  Instead we will choose the lepton doublets to be charged under $SU(2)_B$. Then the lepton Yukawa couplings are identical to the down-type quark Yukawa couplings except that the transpose of $\Sigma$ must be used. Choosing the leptons to transform under $SU(2)_B$ rather than $SU(2)_A$ changes the couplings of the leptons to the heavy gauge bosons which improves the precision fit in \Sec{sec:eft}.  Details of the alternative $SU(2)_A$ lepton charge assignment are given in \App{sec:new-leptons}.

Finally, another deformation of the model would be to couple the down-type quarks and/or leptons to the second Higgs doublet in $\Sigma$, thus going from a type-I to a type-II two Higgs doublet model \cite{Hall:1981bc}.  We will only consider the type-I model here.

\section{Constraints}
\label{sec:pew}

\begin{figure}
\includegraphics[width=\textwidth]{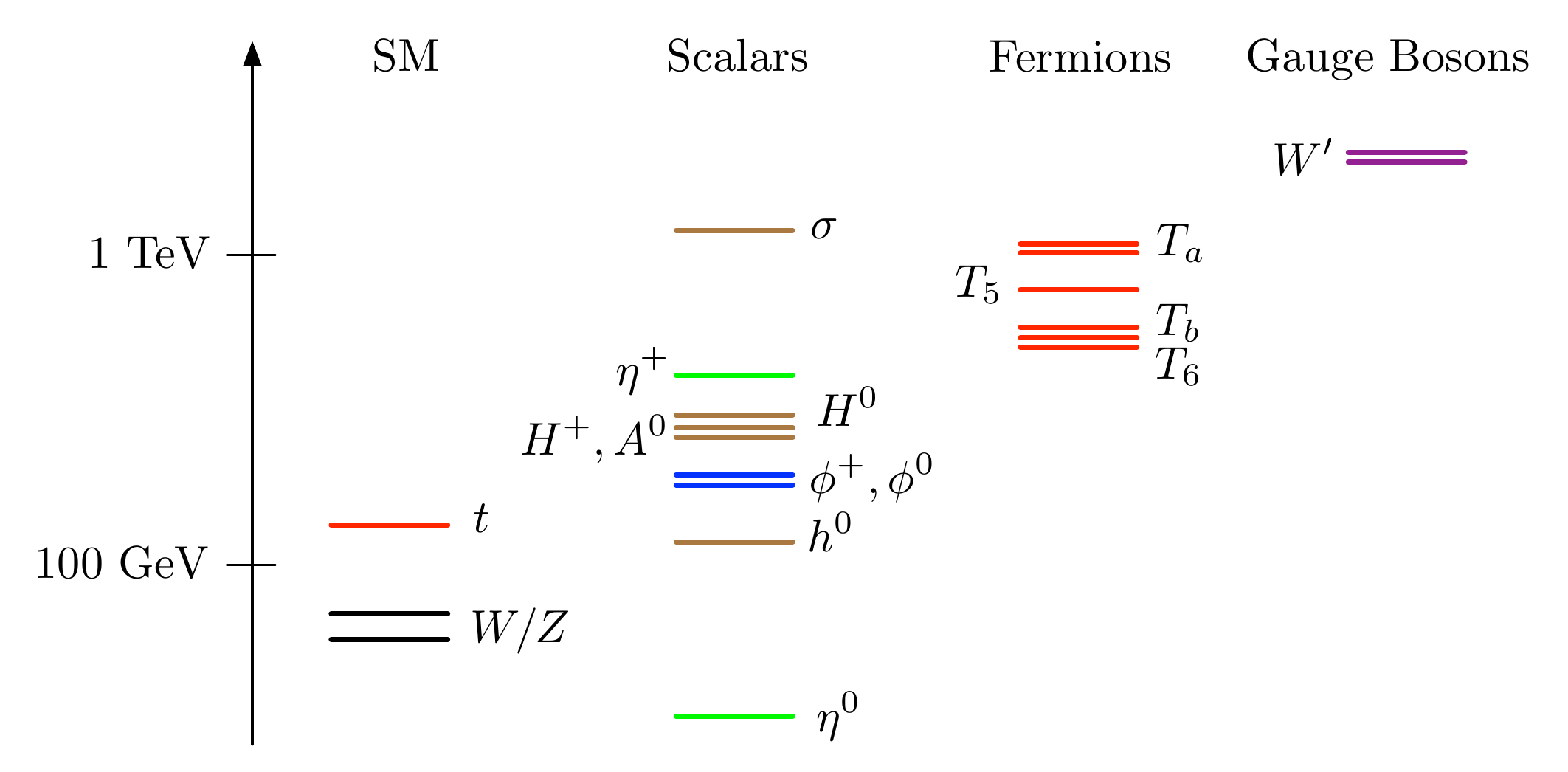}
\caption{An example spectrum in this model.}
\label{fig:spectrum}
\end{figure}

\subsection{Spectrum and Direct Constraints}

We begin by describing the spectrum of new particles which is summarized in \Fig{fig:spectrum}.  The lightest new particle, the $\eta_3$ (hereafter referred to as $\eta^0$), has no gauge interactions and gets its mass only from \Eq{eq:univ_mass}.  Its mass is $m_{4}$, which is a free parameter.  The dominant interactions of $\eta^0$ are to a left- and a right-handed fermion through a Higgs vev, and the interaction strength is proportional to the mass of the fermion over $f$.  Therefore, we require that $m_{4} \gtrsim 10$ GeV so that this particle cannot be seen in low energy flavor factories.  We can determine the higher order couplings of the $\eta^0$ to gauge bosons by inserting spurions that parameterize symmetry breaking loop effects in the kinetic terms.  We find that up to fourth order in scalars there is no coupling to the light gauge bosons, so there will be no constraints from current colliders.  We can also use this method to show that the branching ratio of the Higgses to this particle is neglible. Because it is short lived, there are also no bounds from cosmology or astrophysics.   It may be interesting to search for this particle in precision experiments in the future, particularly in the top sector where it couples most strongly. 

At a mass of about $g_Y f$ which is a few hundred GeV there is the charged $\eta^\pm$ which is an $SU(2)$ singlet and gets its mass from quadratically divergent loops of hypercharge gauge bosons in \Eq{eq:hyp_mass}.  At a similar mass of about $g_{\rm EW} F/4\pi$, there is the electroweak triplet $\phi$ and the five Higgs scalars of the two Higgs doublets which get significant contributions to their masses from loops of gauge bosons shown in \Eq{eq:gauge_mass}. One scalar in the Higgs sector can remain light, particularly in the decoupling limit.  We identify it with the usual SM Higgs and require that its mass be above the the LEP direct search bound of $\sim 114$ GeV  \cite{Barate:2003sz}. By increasing the couplings in the Higgs potential, the Higgs mass can be raised to several hundred GeV. 

Most of the heavier particles in the theory cut off the divergences in the Higgs potential.  At a mass of around $\lambda_0 f\simeq 1$ TeV, there is an additional scalar $\sigma$ with no gauge charges which cuts off divergences from the Higgs quartic.  At a similar mass of order $y_t f$, there are six Dirac fermions with color charge.  There is an $SU(2)$ doublet $T_a$ and a singlet $T_5$ which together cut off the divergence from the top loop.  At a somewhat lower mass there are three associated top partners, another doublet $T_b$ and another singlet $T_6$, which complete the $SO(6)$ multiplet.  The fermions get their mass from the $\Sigma$ vev shown in \Eq{eq:top-partner-mass}. Finally, at a mass around $g_{\rm EW} F$, we have an $SU(2)$ triplet of gauge bosons which cut off quadratic divergences from the light $SU(2)$ gauge boson loops.  

\subsection{Triplet VEV}

One of the most stringent constraints from precision electroweak measurements is that the vev of any electroweak triplet, like our $\phi$, be much smaller than the vev of the Higgses.   A triplet vev can come from either a negative mass squared or a tadpole generated by an operator of the form  $a_\phi \, h^T \phi \, h$, where $a_\phi$ is a dimension 1 coefficient. This is one of the major constraints on other little Higgs models with a triplet that do not utilize $T$-parity, particularly those which use the triplet to generate a collective Higgs quartic such as~\Ref{ArkaniHamed:2002qy}. In our model, the triplet, in the parametrization of \Eq{eq:sig}, does not get a vev at tree level or at one loop. 

To understand this, first note that the contribution to its mass in \Eq{eq:gauge_mass} is positive, and we will assume that the overall triplet mass remains positive. A tadpole is potentially more dangerous. The tree-level Higgs potential in Eqs.~(\ref{eq:quart_field}), (\ref{eq:soft_mass}), and (\ref{eq:Bmu}), does not contain $\phi$ tadpoles so that we can turn to radiative corrections. The fermion couplings as well as the scalar self-interactions preserve a global symmetry for which $\phi$ is the PNGB, so they cannot generate a tadpole either. However, the $SU(2)_A \times SU(2)_B \times U(1)_Y$ gauge interactions break all the shift symmetries protecting $\phi$. To show that they do not generate a tadpole either, we will use a spurion argument using custodial symmetry. The crux of the argument is that $\phi$ transforms as a triplet under custodial symmetry whereas the Higgs vevs are invariant. A tadpole for $\phi$  violates custodial symmetry and can only be generated from custodial symmetry breaking interactions even in the presence of Higgs vevs. The only custodial symmetry violating interaction in the gauge sector is the hypercharge gauge coupling which is proportional to the spurion $T_R^3$. $T_R^3$ is also a triplet under custodial symmetry. Thus we must ask if a term of the form tr$[T_R^3 \phi]$ could be generated after electroweak symmetry breaking. However, the hypercharge gauge coupling also preserves a $T_R^3 \rightarrow - T_R^3$ parity which ensures that any potential from hypercharge interactions involves an even number of $T_R^3$ insertions. But it is easy to see that no non-vanishing operator can be written with an even number of $T_R^3$ spurions and only one field $\phi$. 

Two loop diagrams involving a mixture of gauge, Higgs, or fermion interactions are expected to generate a tadpole for $\phi$ with coefficient
\be
a_\phi \sim \frac{f}{(16\pi^2)^2} 
\log^2\left(\frac{\Lambda^2}{f^2}\right) \ .
\ee
This gives a vev for $\phi$ which is smaller by $1/16 \pi^2$ than the bounds from precision measurements.

\subsection{Effective Dimension Six Operators}
\label{sec:eft}

The remaining constraints on the model come from precision measurements at scales well below the mass of the new particles, so we can use an effective field theory approach.  Furthermore, we focus on tree-level effects; loop effects are suppressed by the mass of the new heavy particle as well as by $16\pi^2$ and are small enough to not significantly affect precision electroweak observables.  Physics at or above the cutoff of our theory is expected to generate a slew of higher-dimensional operators suppressed by $\Lambda $.  For $\Lambda \simeq 10$ TeV these operators are too small to contribute to electroweak measurements, but we must assume that the UV physics does not generate operators which strongly violate flavor, $CP$, or baryon/lepton number.  

We are now ready to consider the tree-level effects from integrating out heavy gauge bosons, fermions, and scalars.  Using the notation of \Refs{Han:2004az, Han:2005dz}, the dimension six operators which are highly constrained and may be generated at tree level in our model are:
\begin{eqnarray}
   O_{W\!B}&=&( h^\dagger \tau^a h) W^a_{\mu\nu} B^{\mu\nu},\quad\quad\quad\quad\ \,
   O_h=\ | h^\dagger D_\mu h|^2,\nonumber\\
  O_{hf}^s& =& i (h^\dagger D^\mu h)(\overline{f} \gamma_\mu f) + {\rm h.c.}, \ \ \ \ \,
  O_{hf}^t =\ i (h^\dagger \tau^a D^\mu h)(\overline{f} \gamma_\mu \tau^a f)+
     {\rm h.c.},\nonumber\\
 O_{ff'}^s&=&\frac{1}{1+\delta_{ff'}} (\overline{f} \gamma^\mu f) (\overline{f'} \gamma_\mu f'), \ \ \
 O_{ff'}^t=\ \frac{1}{1+\delta_{ff'}} (\overline{f} \gamma^\mu \tau^a f)  
                 (\overline{f'} \gamma_\mu \tau^a f') \ .
\label{eq:operators}
\end{eqnarray}
Here $h$ is either one of the Higgs doublets, $f, f' = q, u^c, d^c, l, e^c$ are the SM fermions, and $W_{\mu\nu}^a$ and $B^{\mu\nu}$ are the field strength tensors of $SU(2)_L$ and $U(1)_Y$.  The operators with a superscript $t$ have $SU(2)_L$ triplet contractions and can only involve the $SU(2)$ doublets, $q$ or $l$.  The singlet operators with superscript $s$ can exist for any type of fermion. The operator $O_{W\!B}$ is equivalent to the oblique $S$ parameter and is potentially dangerous because it cannot be forbidden by symmetries. However, it is not generated at tree level in our model, and we will not consider it further.  $O_h$ parameterizes custodial symmetry breaking and corresponds to the oblique $T$ parameter. Because our sigma model preserves custodial symmetry, we do not have a large tree level contribution to $T$ from the sigma model. The $SU(2)_L$ gauge boson partners and the heavy scalars preserve custodial $SU(2)$ as well, thus integrating them out cannot generate $T$ either. Finally, tree level exchanges of fermions cannot give rise to operators without fermions. Thus both $S$ and $T$ vanish at tree level in our model.\footnote{A tree-level $T$ would have been generated if we had introduced a partner for the hypercharge gauge boson.}

To understand which of the remaining operators---all involving fermions---are generated we consider integrating out fermion, scalar and gauge partners in turn. Only the top and left-handed bottom have fermionic partners. Integrating them out leads to the operators in \Eq{eq:operators} involving the third generation which are not very constrained. A notable exception is $Z \rightarrow  b \bar{b}$. However, as we already discussed in \Sec{sec:light_quarks}, the bottom quark does not mix with a heavy  fermion partner with different gauge charges, and therefore its coupling to the $Z$ is unchanged. Integrating out scalars could potentially generate dimension 6 operators. However our scalar partners only couple very weakly (proportional to Yukawa couplings) to the light fermions and therefore any operators with fermions which might be generated are too small to be relevant. 

Thus the only precision electroweak constraints come from tree-level exchanges of heavy gauge bosons.  The mass of these bosons is controlled by $F$. Therefore precision electroweak will not constrain $f$, and we may choose $f \simeq 1$ TeV for which the quartic and fermion sectors do not generate any fine-tuning.  The masses of the fermion and scalar partners will also be around 1 TeV. This is a novel feature of our model compared with all other little Higgs models which typically have much heavier fermion partners. In our model, this stems from the decoupling of the gauge and fermion masses by controlling them with separate scales.

The heavy gauge bosons, $A_H^a$, couple to the light left-handed quark doublets with a coupling of 
\begin{equation}
-\frac{g_A^2}{ \sqrt{g_A^2 + g_B^2}}A_{H\mu}^a\, \bar{q}_L \sigma^\mu \frac{\tau^a}{2} q_L+ {\rm h.c.} \, , 
\label{eq:quark_Wp}
\end{equation}
where $\tau^a$ are the Pauli matrices which contract gauge indices, and we have suppressed flavor indices which are contracted in the usual way. As discussed in \Sec{sec:light_quarks}, the lepton doublet is charged under $SO(6)_B$, so it has a different coupling to the heavy gauge bosons:
\begin{equation}
\frac{g_B^2}{ \sqrt{g_A^2 + g_B^2}}A_{H\mu}^a\, \bar{\ell}_L \sigma^\mu \frac{\tau^a}{2} \ell_L+ {\rm h.c.} \ .
\end{equation}
 The heavy gauge bosons also couple to the Higgs:
\begin{equation}
\frac{i}{2} \frac{g_A^2 - g_B^2}{ \sqrt{g_A^2 + g_B^2}} {\rm tr} \left( 
A^{(6)}_{H\mu} [\Pi_h, D^\mu \Pi_h]
\right),
\label{eq:higgs_gauge_partners}
\end{equation}
where $\Pi_h$ is given in \Eq{eq:pih} and $D^\mu$ is a covariant derivative with respect to SM gauge fields only.  Note that this coupling vanishes in the ``$T$-parity limit" $g_A=g_B$. Because we have not measured the properties of physical Higgs bosons, the only bounds on operators with the Higgs come from inserting the Higgs vev.  When plugging in the Higgs vevs they always appear in the combination $v_{\rm EW}^2=v_1^2+v_2^2$, therefore the constraints on our two-Higgs doublet scenario are identical to those for a model with a single Higgs field.

Integrating out the heavy triplet of gauge bosons and denoting the coefficients of the operators given in \Eq{eq:operators} by $a$, we get
\begin{align}
a^t_{ql} &= \frac{ g_A^2 \, g_B^2}{4 m_{W'}^2(g_A^2 + g_B^2)} = 
                  \frac{g_{\rm EW}^2}{4 m_{W'}^2} ,
&a^t_{ll} &= - \frac{ g_B^4}{4 m_{W'}^2(g_A^2 + g_B^2)} = 
               	   - \frac{g_{\rm EW}^2}{4 m_{W'}^2} \, \cot^2\! \theta_g, \nonumber \\
a^t_{hq} &=  -\frac{ g_A^2\, (g_A^2 - g_B^2) }{4 m_{W'}^2 (g_A^2 + g_B^2)} = 
               	    \frac{g_{\rm EW}^2}{4 m_{W'}^2}  (1\!-\! \tan^2\! \theta_g) ,
&a^t_{hl} &= \frac{ g_B^2\, (g_A^2 - g_B^2) }{4 m_{W'}^2 (g_A^2 + g_B^2)}= 
               	    \frac{g_{\rm EW}^2}{4 m_{W'}^2}  (1\!-\!\cot^2\! \theta_g) ,
\end{align}
where we have omitted $a^t_{qq}$ because it is poorly constrained, and the mass of the heavy gauge boson is given in \Eq{eq:heavy_gauge}.  These are all the operators in \Eq{eq:operators} generated by exchange of heavy gauge bosons.  Using the results of~\Ref{Han:2004az}, we can compute the effect of these higher dimensional operators on all measurements and determine a bound.  We require that the difference in $\chi^2$ between our model and the SM is less than 4 (for $95\%$ confidence) or 9 (for $3\sigma$)  and use this to place a bound on the mass of the heavy gauge bosons.  We use the SM $SU(2)$ gauge coupling $g_{\rm EW} = 2 M_W / v \simeq 0.65$ as an input, so our fit is a function of two model parameters, $\tan \theta_g = g_A/g_B$ and the mass of the heavy gauge boson $m_{W'}$.  The electroweak fit also depends on loop corrections from the Higgs sector. Working in the decoupling limit, we will only include the usual logarithmic dependence of $S$ and $T$ on the mass of the lightest Higgs, $M_{h^0}$, as detailed in Appendix A of~\Ref{Han:2004az}.  For the Higgs mass, we show two benchmark points, 125 and 250 GeV.  

\begin{figure}
\includegraphics[scale=0.6]{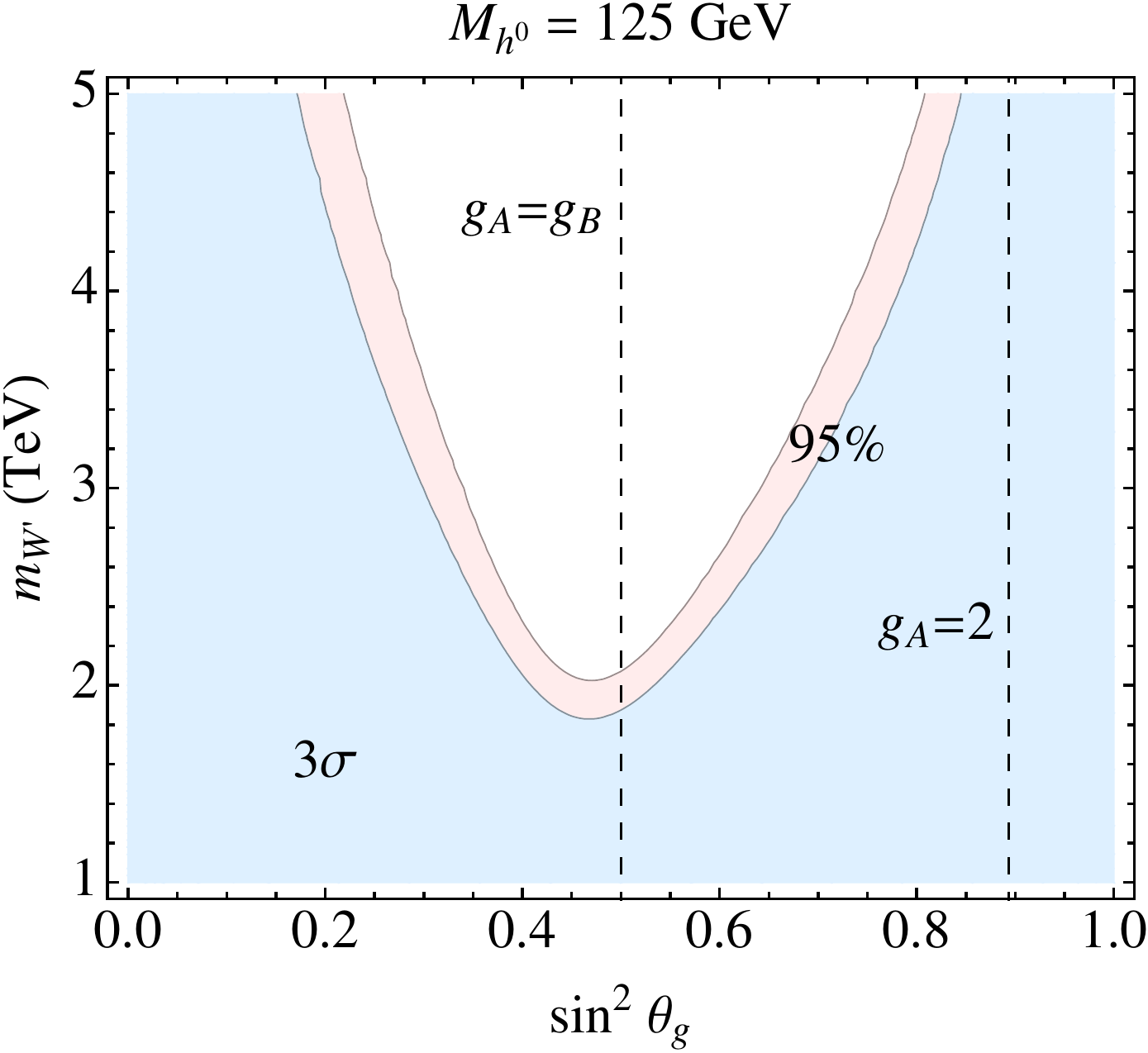}
\includegraphics[scale=0.6]{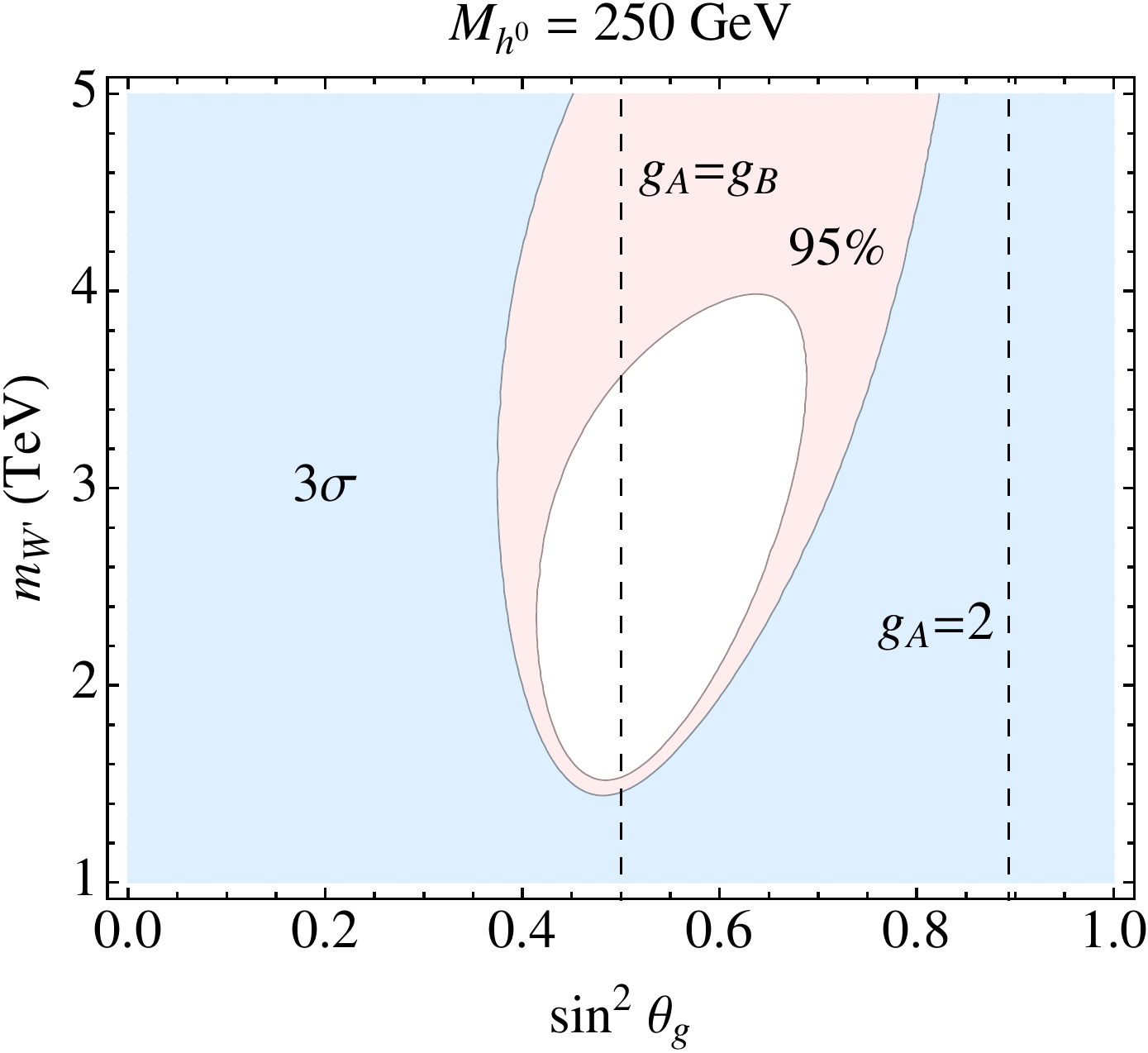}
\caption{The allowed masses for the heavy gauge boson $m_{W'}$ as a function of $\sin^2 \theta_g$, with $M_{h^0} = 125$ (250) GeV on the left (right).  The light blue region is excluded at $3\sigma$ with respect to the SM, the lavender region is disfavored at 95\%, and the white region is allowed.  We have also marked where $g_A=g_B$ and $g_A = 2$ using \Eq{eq:gauge_coupling}. }
\label{fig:electroweak-fit}
\end{figure}

The results of the electroweak fit are shown in \Fig{fig:electroweak-fit}.  We see that a large region of parameter space is open as long as $g_A \simeq g_B$ where the gauge boson partners decouple from the Higgses (``$T$-parity limit").  For a Higgs mass of 125 GeV, the allowed region gets quite large as we increase $m_{W'}$ which makes sense because that is the SM limit.  On the other hand, increasing $m_{W'}$ makes the allowed region smaller for heavier Higgs masses which reproduces the well-known fact that a heavy Higgs is disfavored in the SM.  From \Fig{fig:electroweak-fit} we see that the gauge partner can be quite light, lighter than 2 TeV for a heavy Higgs.  Finally, it is interesting to note that much of the allowed parameter region, including our benchmark of $M_{h^0}=250$ GeV, $m_{W'} = 2$ TeV, and $g_A=g_B$, fits the electroweak data better than the SM does.  While we take this as our benchmark point, we will show in \Sec{sec:collider} that the collider phenomenology becomes more interesting if the two gauge couplings are not exactly equal.  Therefore, we will imagine that the two couplings are similar but not equal.  

We are now in position to calculate the fine-tuning in the $SU(2)$ gauge sector. Using Eqs.~(\ref{eq:fine_tuning}) and~(\ref{eq:gauge_mass}), we find 
\be
\label{eq:gauge_finetuning}
\Psi_g = \frac{9\, g_{\rm EW}^2 \sin^2\beta \,m_{W'}^2}{32 \, \pi^2 M_{h^0}^2 }
               \log \left( \frac{\Lambda^2}{m_{W'}^2} \right).
\ee
which becomes $\Psi_g \simeq 2$ for our benchmark point, comparable to the other ``tunings" in the model.

\section{Collider Phenomenology}
\label{sec:collider}

This model has a cornucopia of new particles whose spectrum is described at the beginning of \Sec{sec:pew} and shown in \Fig{fig:spectrum}.  The main phenomenological features that distinguish this model from traditional little Higgs scenarios are the particularly light top partners and the presence of the uneaten triplet scalars $\phi$.  In this section, we discuss the collider phenomenology relevant for discovering these new modes.

\begin{table}
\begin{center}
\begin{tabular}{c|c|c|c|c|c}
 &  $U(1)_{\rm EM}$ & Mass & SM Mixing &  Decay to SM  & Cascade Decay\\
 \hline
$T_a^{(u)}$ & $\phantom{-}2/3$ & $\sqrt{|y_1|^2 + |y_2|^2}f$ 
          & \checkmark & $X^0 t_R$ & $X^0 T_6$ \\
$T_a^{(d)}$ & $-1/3$ & $\sqrt{|y_1|^2 + |y_2|^2}f$ 
           & \checkmark & $X^- t_R$ &  $X^- T_6$\\ 
$T_b^{(5/3)}$ & $\phantom{-}5/3$ & $|y_1|f$ & & $X^+ t_R$ & \\
$T_b^{(2/3)}$ & $\phantom{-}2/3$  & $|y_1|f$ & & $X^0 t_R$ & \\
$T_5$ & $\phantom{-}2/3$ & $\sqrt{|y_1|^2 + |y_3|^2}f$ 
           & \checkmark & $X^0 t_L$, $X^+ b_L$ & $X^- T_b^{(5/3)}$, $X^0 T_b^{(2/3)}$  \\
$T_6$ & $\phantom{-}2/3$  & $|y_1|f$ & & $X^0 t_L$, $X^+ b_L$ &
\end{tabular}
\end{center}
\caption{The properties of the top partners.  The second column gives the electromagnetic charge of the Dirac multiplet, and the third column is the mass up to $\mathcal{O}(v^2/f^2)$ corrections.  The fourth column denotes whether the particle mixes with the SM particle with the same quantum numbers and therefore cuts off quadratic divergences.  The fifth column lists the dominant decay channels to SM fermions.  Here, $t_L$ and $b_L$ are the left-handed top and bottom quarks, while $t_R$ is the right-handed top quark.  $X^0$ refers to a neutral Higgs $h^0$/$H^0$/$A^0$ or a longitudinal gauge boson $Z^0_L$, and $X^\pm$ refers to a charged Higgs $H^\pm$ or a longitudinal gauge boson $W^\pm$.  The last column indicates the leading cascade decay from a ``heavy'' top partner to a ``light'' top partner, omitting possible transitions $T_a \rightarrow T_5$ or vice-versa.  While the PNGBs $\eta$ and $\phi$ can be produced in decays of the top partners, those branching fractions are either suppressed by $(v_{\rm EW}/f)^2$ or, in the case of $\eta^\pm$, by phase space.}
\label{table:top-partners}
\end{table}

With such light top partners, the collider phenomenology is dominated by production and decay of the new colored states.  As shown in \Tab{table:top-partners}, there are six Dirac fermions with SM color charge and masses of order $y_t f$.  The ``heavy'' top partners $T_a$ and $T_5$ are responsible for regulating the Higgs potential.  The ``light'' top partners $T_b$ and $T_6$ play no role in cutting off divergences in the Higgs potential; they simply appear in the theory because of the $SO(6)$ symmetry structure.  

All the top partners can be pair-produced through QCD processes, with cross sections around 1 pb for 600 GeV top partners at a 14 TeV LHC.  Previous studies of $\sim2$ TeV top partners in traditional little Higgs theories have emphasized the important role of single production of top partners through $Wb$ fusion \cite{Han:2003wu,Perelstein:2003wd}, and, while this process still has a large cross section for the $T_a^{(u)}$, $T_5$, and $T_6$ states \cite{Willenbrock:1986cr}, it no longer overwhelms QCD pair production.\footnote{In addition, the state $T_a^{(d)}$ mixes with the SM left-handed bottom so it can be produced in association with a top quark through an off-shell $W$ boson, but this mode also does not dominate over QCD pair production.}  We will focus on strong QCD pair production since it gives more striking collider signatures compared to weak single production.

\begin{figure}
\begin{center}
\includegraphics[width=0.5\textwidth]{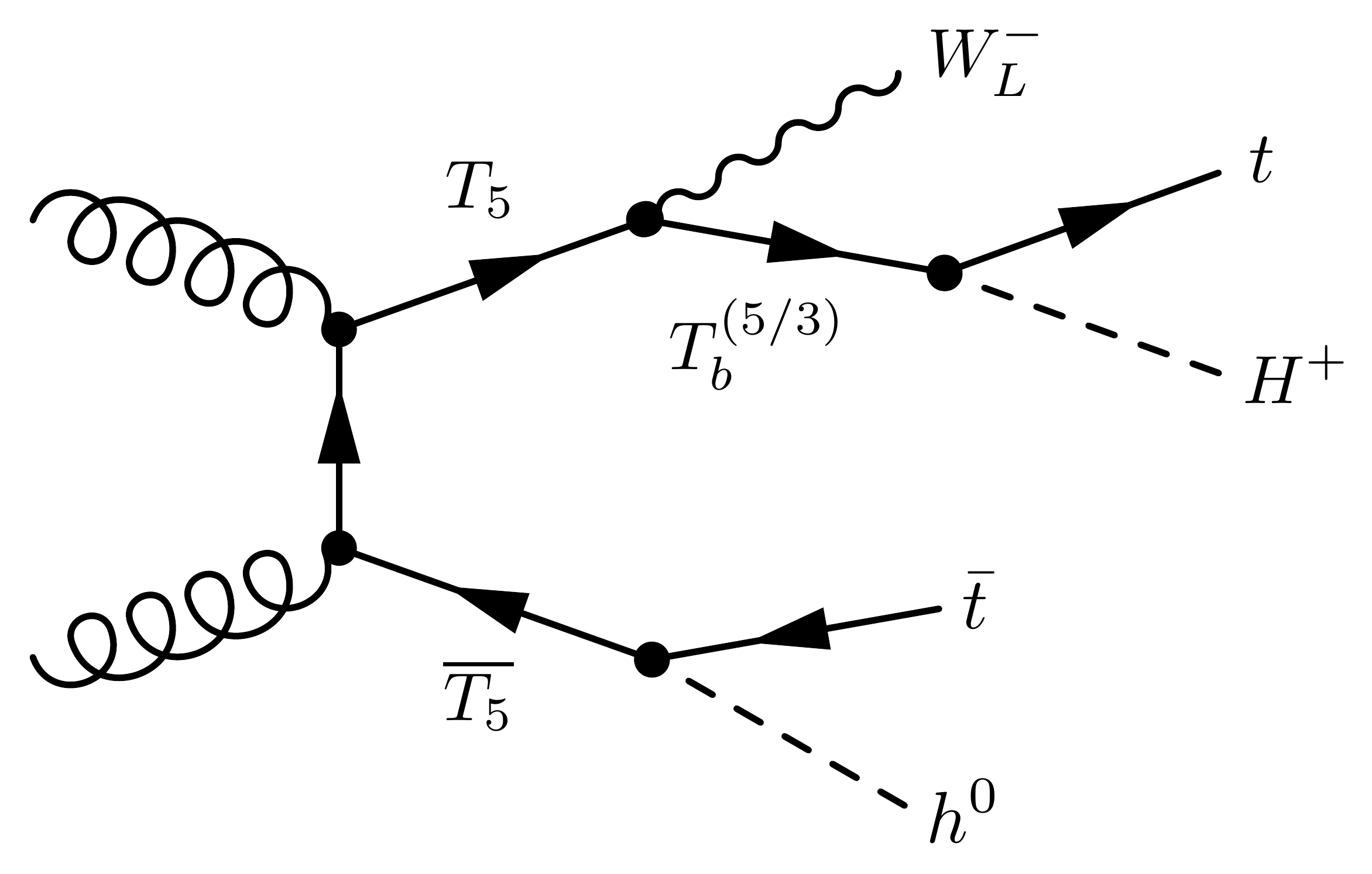}
\end{center}
\caption{A representative cascade decay from pair production of top partners.  Top partners dominantly decay directly to a third generation quark and a Higgs/gauge boson, here shown by the $\overline{T_5}$ decay.  If phase space is available, a ``heavy'' top partner can also cascade decay through a ``light'' top partner, shown here by $T_5 \rightarrow T_b^{(5/3)}$.  The final state top quarks and gauge/Higgs bosons then undergo their usual decays.}
\label{fig:cascade}
\end{figure}

When pair produced, each top partner will decay to third generation fermions.  The largest coupling in the theory is the top Yukawa coupling from \Eq{eq:top_yukawa}, so two-body decays mediated by those operators will dominate over other decay channels.  The leading decay modes are detailed in \Tab{table:top-partners}, and involve either a Higgs boson ($h/H/A^0/H^\pm$) or the longitudinal component of a $W/Z$ boson (as expected from the Goldstone equivalence theorem). In addition, if phase space allows, a ``heavy'' top partner can cascade decay to a ``light'' top partner and a Higgs/gauge boson.  Thus, the final state will typically contain two third generation fermions and between two and four Higgs/gauge bosons, as shown in \Fig{fig:cascade}.  The Higgses themselves decay similar to a usual two Higgs doublet model \cite{Gunion:1989we}, so the final state will have many third generation fermions and electroweak gauge bosons.

Because we are taking $F > f$, the gauge partners in this model are parametrically heavier than the top partners.  That said, the bound on $m_{W'}$ is similar to traditional little Higgs models, and therefore the expected collider phenomenology in the gauge boson sector is similar to previous studies such as \Ref{Han:2003wu}.  The production cross section for the gauge bosons is quite large, about 1 pb for a mass of 2 TeV at a 14 TeV LHC.  Furthermore, these particles have substantial coupling to leptons and will decay to them a non-trivial fraction of the time, making their discovery quite feasible.  If kinematically allowed, the heavy gauge bosons can also decay to pair of top partners (or to a third generation quark and a top partner), leading to the same kinds of final states as discussed in \Fig{fig:cascade}. 

One major difference between our model and traditional little Higgs phenomenology is
that the gauge partners can also decay to a pair of triplets $\phi$, since these states are no longer eaten by the $SU(2)_L$ gauge boson partners.  The coupling of $\phi$ to the heavy gauge bosons is similar to the coupling of the Higgs given in \Eq{eq:higgs_gauge_partners}. While this coupling vanishes in the ``$T$-parity limit'' where $g_A = g_B$, it is generically present, albeit suppressed compared to the couplings to SM fermions.  This is likely the most promising method for discovery of the triplet $\phi$, since its electroweak production cross section is somewhat small, and decay of top partners into these particles is suppressed by $(v_{\rm EW}/f)^2$.\footnote{This suppression does not appear in the alternative top Yukawa structures shown in \App{sec:alttop}.}   When produced, $\phi$ will decay primarily through the top Yukawa couplings to pairs of third generation quarks.\footnote{The $\phi$ scalar could decay via the gauge or kinetic interactions of \Eq{eq:kinetic}, but by expanding out the parameterization in \Eq{eq:sig}, one can see that these decays are suppressed.}  Therefore events where one heavy gauge boson decays to a pair of $\phi$'s will have four third generation quarks in the final state and could also be quite spectacular.   If the LHC discovers gauge bosons which are heavier than the top partners---and if it also discovers scalars which could have been eaten by these gauge bosons---then this would be strong evidence for the modular gauge sector proposed in this paper.  

Finally, there may be lower energy manifestations of this model.  As discussed in \Sec{sec:pew}, the mass of the neutral $\eta^0$ is a free parameter, and it can be as light as 10 GeV.  Its coupling to top quarks is $\mathcal{O}(m_t/f)$ and so it may be radiated off in top pair production events.  When produced, $\eta^0$ will decay dominantly to $b \bar{b}$, yielding $t\bar{t}b \bar{b}$ events that might be visible even at the Tevatron.

\section{Conclusions}
\label{sec:conc}

While little Higgs models provide an appealing solution to the hierarchy problem, concrete models in the literature suffer from two generic problems.  The first is that it is difficult to generate a quartic coupling for the Higgs while not violating custodial symmetry.  The second is that most models have at least 10\% fine-tuning from the top sector owing to constraints from precision electroweak observables.  In this paper, we have presented a model that solves both of these problems.  

Our strategy was to take a modular approach to model building, optimizing the structure of the quartic, gauge, and fermion sectors separately.  For the Higgs quartic coupling, we began by building a simple non-linear sigma model that contained two Higgs doublets as PNGBs, using the symmetry breaking structure $SO(6)\times SO(6) \rightarrow SO(6)$.  The upper $SO(4)$ of the $SO(6)$'s is identified with the SM $SU(2)_L\times SU(2)_R$ ensuring an approximate custodial symmetry.  A collective quartic coupling for the Higgs is generated with the help of the electroweak singlet, but this is not a dangerous singlet \cite{Schmaltz:2008vd} because it transforms under a global symmetry. 

For the gauge sector, we employed a modular gauge structure in order to decouple the mass of the gauge partners from the Higgs and top partners.   We introduced a new non-linear sigma field which is a singlet under all the global symmetries which protect the Higgs, but which transforms under the same gauge symmetries.  This new sector has a decay constant $F$ that is higher than the decay constant for the Higgses $f$.  In this way, we can raise the mass of the new gauge bosons to a few TeV, above the bound from precision electroweak measurements, without introducing fine-tuning in the top and Higgs sectors.  This method for implementing collective gauge couplings is completely generic, and can be used as a model building tool for any little Higgs model.   In our case, the phenomenological implication of this modular gauge structure was the appearance of a set of uneaten PNGB modes which fill out a \textbf{6} of $SO(4)$.

Finally, for the top Yukawa, we implemented a collective symmetry breaking structure that minimized the top partner masses for a fixed decay constant $f$.  The sensitivity of the Higgs potential to the top Yukawa is dramatically decreased by a factor of 9 compared to most little Higgs theories, making for considerably more natural electroweak symmetry breaking.  Correspondingly, our 500-800 GeV top partners are considerably lighter than in other little Higgs constructions without $T$-parity, leading to a more optimistic collider phenomenology.  Production and decay of the top partners leads to events with a large number of third generation fermions and SM gauge bosons.  As the LHC has already started to take data, looking for such events will soon become very interesting.  

\section*{Acknowledgements}

We thank Zackaria Chacko, Andy Cohen, Liam Fitzpatrick, Zhenyu Han, Arvind Rajaraman, Christian Spethmann, and Jay Wacker for helpful comments.  D.S. is supported by the National Science Foundation, M.S. is supported by the U.S. Department of Energy under DE-FG02-01ER-40676, and J.T. is supported by the U.S. Department of Energy (D.O.E.) under cooperative research agreement DE-FG02-05ER-41360.

\appendix

\section{Group Theory Results}
\label{sec:grouptheory}

An $SO(4)$ group can be written as the product of $SU(2)_L \times SU(2)_R$ groups.  In terms of the canonical $SO(4)$ generators, which are imaginary anti-symmetric matrices, the $SU(2)$ generators are given as follows:
\begin{equation}
T_L^1 =   \frac{i}{2}
  \left( \begin{array}{cccc}
    0 & 0 & 0 & 1 \\
    0 & 0 & 1 & 0 \\
    0 & -1 & 0 & 0 \\
    -1 & 0 & 0 & 0
  \end{array} \right),  \quad
T_L^2 =   \frac{i}{2}
  \left( \begin{array}{cccc}
    0 & 0 & 1 & 0 \\
    0 & 0 & 0 & -1 \\
    -1 & 0 & 0 & 0 \\
    0 & 1 & 0 & 0 
  \end{array} \right), \quad
T_L^3 =   \frac{i}{2}
  \left( \begin{array}{cccc}
    0 & 1 & 0 & 0  \\
    - 1 & 0 & 0 & 0 \\
    0 & 0 & 0 & 1 \\
    0 & 0 & -1 & 0
  \end{array} \right),
  \label{eq:T_L}
\end{equation}
\begin{equation}
T_R^1 =   \frac{i}{2}
  \left( \begin{array}{cccc}
    0 & 0 & 0 & -1 \\
    0 & 0 & 1 & 0 \\
    0 & -1 & 0 & 0 \\
    1 & 0 & 0 & 0
  \end{array} \right),  \quad
T_R^2 =   \frac{i}{2}
  \left( \begin{array}{cccc}
    0 & 0 & -1 & 0 \\
    0 & 0 & 0 & -1 \\
    1 & 0 & 0 & 0 \\
    0 & 1 & 0 & 0 
  \end{array} \right), \quad
T_R^3 =   \frac{i}{2}
  \left( \begin{array}{cccc}
    0 & -1 & 0 & 0  \\
    1 & 0 & 0 & 0 \\
    0 & 0 & 0 & 1 \\
    0 & 0 & -1 & 0
  \end{array} \right).
  \label{eq:T_R}
\end{equation}
These are normalized so that the usual $SU(2)$ commutation relation, $[T^a,T^b]=i \, \epsilon_{abc} T^c$ holds.  We normalize the rest of the $SO(6)$ generators so that all of them have ${\rm Tr}\left( T_a T_b \right) = \delta_{ab}$.

A $\bf{4}$ of $SO(4)$ can be written as a complex $SU(2)$ doublet in the following way:
\be
H=\frac{1}{\sqrt{2}} \left( \begin{array}{c}  h_3 + i h_4 \\ h_1 - i h_2 \end{array} \right).
\label{eq:higgsdoublet}
\ee
This means that to contract the $SU(2)$ indices in $\Delta$ with those in $\Sigma$ which are in $SO(4)$ notation, as is done in \Eq{eq:univ_mass}, we need a $2\times 6$ matrix which encodes this information.  This matrix is given by
\begin{equation}
M_{26} =\frac{1}{2} \left( \begin{array}{cccccc}
    0 & 0 & 1 & i & 0 & 0  \\
    1 & -i & 0 & 0 & 0 & 0 
  \end{array} \right),
\end{equation}
where the normalization is chosen so that the mass terms in \Eq{eq:univ_mass} have the same magnitude as \Eq{eq:soft_mass}.  

The algebra of $SO(6)$ is the same as that of $SU(4)$, and since many readers are more familiar with the $SU(N)$ groups, we here include how to embed the goldstone bosons of our non-linear sigma model in $SU(4)$ language.  To write $\Pi$ from \Eq{eq:pi},
\be
\Pi = 
\frac{1}{2}
  \left( \begin{array}{cc}
    \phi_a \tau^a +\frac{\sigma}{\sqrt{2}} \one & 0  \\
    0 & \eta_a \tau^a - \frac{\sigma}{\sqrt{2}} \one \\
  \end{array} \right),
\ee
where each entry represents a $2\times2$ block and $\tau^a$ are the Pauli matrices.  To write $\Pi_h$ from \Eq{eq:pih} in $SU(4)$ language
\be
\Pi_h = 
\frac{1}{2}
  \left( \begin{array}{ccc}
    0 & \tilde{H_2^*}-i \tilde{H_1^*} & H_2-i H_1   \\
    \tilde{H_2}^T+i \tilde{H_1}^T & 0 & 0 \\
    H_2^\dagger + i H_1^\dagger & 0 & 0
  \end{array} \right),
\ee
where the first row and column are length two, $H$ is a complex doublet in the notation of~\Eq{eq:higgsdoublet}, and $\tilde{H} = i \tau^2 H$.

\section{Coleman--Weinberg Calculations}
\label{sec:cw}

In order to calculate the one loop effects from the different operators in the model, we compute the Coleman--Weinberg potential~\cite{Coleman:1973jx} which is given by 
\begin{equation}
V_{\rm CW} = \frac{\Lambda^2}{32 \pi^2}{\rm Str}\left( M^2(\Sigma) \right) + 
                   \frac{1}{64 \pi^2} {\rm Str}\left[ M^4(\Sigma)\left( \log \left(\frac{M^2(\Sigma)}{\Lambda^2} \right)- \frac{1}{2} \right) \right],
\label{eq:Vcw}
\end{equation}
where $M^2(\Sigma)$ is the mass matrix in a $\Sigma$ background of the field that is being traced over, and $\Lambda$ is the cutoff of the theory.   We take $\Lambda \simeq 4 \pi f \simeq 10 \, {\rm TeV} $.  We rewrite the second term as follows
\be
-\frac{1}{64\pi^2}  {\rm Str}\left[ M^4(\Sigma) \log \left( \frac{\Lambda^2}{\mu^2} \right) 
+  M^4(\Sigma) \log \left( \frac{\mu^2}{M^2(\Sigma) }\right)  +\frac{1}{2}M^4(\Sigma)  \right],
\ee
where we choose $\mu$ to minimize the finite pieces and higher loop corrections.  This $\mu$ turns out to be the mass of the particle which cuts of the divergence in the sector we are computing. 

\subsection{Quartic}
\label{sec:CWhiggs}

To parameterize the fluctuations about a $\Sigma$ background in the scalar sector we take 
\begin{equation}
\Sigma = \langle \Sigma \rangle \, \delta \Sigma \ ,
\end{equation}
where $\delta \Sigma$ are the fluctuations.  This parameterization ensures that the kinetic terms of the fluctuation fields are canonically normalized.   We also restrict ourselves to the operators in \Eq{eq:V_quart} because the coefficients of the operators in \Eqs{eq:soft_mass}{eq:Bmu} are small, so their radiative corrections will be negligible.  
The quadratically divergent correction to the potential is 
\begin{equation}
 -\frac{3 f^2 \Lambda^2}{16 \pi^2}  \left( \lambda_{65} \left| \Sigma_{65} \right|^2 
     + \lambda_{56} \left| \Sigma_{56} \right|^2 \right) ,
\end{equation}
which is the same size but opposite sign as the tree level piece, so it does not generate any dangerous operators.  The value of the quartic is dependent on the UV physics, but it is not fine-tuned.  This confirms that $\sigma$, which cuts off the divergence in the quartic sector, is not a dangerous singlet because no tadpole for it is generated at the one loop quadratic divergent level.  The only tadpole comes from the ``soft'' $B_\mu$ term in \Eq{eq:Bmu}.

The logarithmically divergent piece of the potential contains the operators in the tree level potential, but it also adds new operators given by
\begin{equation}
\frac{\lambda_{65} \lambda_{56}}{16 \pi^2} \log \left( \frac{\Lambda^2}{m_\sigma^2} \right)
     \left[ f^2 \left( h_1^T h_1 + h_2^T h_2  \right)  -
          \frac{2}{3}\left( h_1^T h_1 \right)^2 -  \frac{2}{3}\left(h_2^T h_2 \right)^2 -
         2 \left(h_1^T h_1 \right)  \left(h_2^T h_2 \right) \right],
\end{equation}
and we have ignored operators of dimension 5 and higher suppressed by powers of $1/f$.  From this, one can see that the new quartics generated are numerically quite small, so ignoring them in the analysis of \Sec{sec:scal_pot} is justified.

\subsection{Gauge}
\label{sec:CWgauge}

Loops of heavy and light gauge bosons also generate corrections to the scalar potential.  Because of collective breaking, there is no quadratically divergent potential generated at one loop.  The logarithmically divergent mass terms generated are given in \Eq{eq:gauge_mass}, and the new interaction terms generated are 
\begin{eqnarray}
-\frac{3 \, g_A^2 \, g_B^2 \, F^2}{64 \pi^2 f^2} \log \left( \frac{\Lambda^2}{m_{W'}^2} \right)  
\Bigg[ \frac{1}{16} \left(h_1^T h_1 + h_2^T h_2 \right)^2   
       + \frac{1}{2}\phi_a \phi^a \left( h_1^T h_1  + h_2^T h_2  \right) + 
       \frac{1}{3} (\phi_a \phi^a )^2 \Bigg] ,
\end{eqnarray}
where we have ignored corrections of order $f^2/F^2$.  The fact that the quartics are negative makes sense because the potential is a periodic function of the field, so if the second derivative is positive, then the fourth derivative should be negative. These quartics are again all quite small so they do not destabilize the vacuum calculated in \Sec{sec:scal_pot}.  After electroweak symmetry breaking, these loops will generate a negative mass for the triplet $\phi$, but this mass will be parametrically smaller than the one generated by loops shown in \Eq{eq:gauge_mass}.

\subsection{Hypercharge}
\label{sec:CWhyp}

Because hypercharge is not implemented collectively, loops of the SM hypercharge gauge boson generate the quadratically divergent mass terms given in \Eq{eq:hyp_mass} as well as the following quartics:
\begin{equation}
-\frac{3 g_Y^2 \Lambda^2}{32 \pi^2 f^2} \left[ 
 \frac{1}{3}(\eta_1^2 + \eta_2^2 )^2 + \frac{1}{3}(\eta_1^2 + \eta_2^2 ) \, \eta_3^2 
 +\frac{1}{2}(\eta_1^2 + \eta_2^2 )(h_1^T h_1 +h_2^T h_2) + {\rm Higgs\, quartics}
\right] ,
\end{equation}
where the Higgs quartics can be written as follows
\begin{equation}
8 \,{\rm tr}\left(\frac{1}{12}\Pi_h^4 (T^3_R)^2 +  \frac{1}{4} \Pi_h^2 \, T^3_R \, \Pi_h^2 \, T^3_R
                + \frac{1}{3}\Pi_h \, T^3_R \, \Pi_h^3 \, T^3_R \right).
\label{eq:hypercharge_quartics}
\end{equation}
Explicit calculation shows that these interaction terms do not destabilize the Higgs potential analyzed in \Sec{sec:scal_pot}, and that they are small so they do not change the vevs by very much.  Some of the terms in \Eq{eq:hypercharge_quartics} are not custodially symmetric, but they are proportional to hypercharge and similar to the terms which would be present in any two Higgs doublet model, so they do not conflict with precision electroweak constraints.  There will also be corrections to the potential which are logarithmically divergent and proportional to $g_Y^4$ or $g_{\rm EW}^2 g_Y^2$, but these are even smaller and can be neglected.

\subsection{Fermions}
\label{sec:CWfermions}

The only significant radiative correction from loops of fermions comes from the top Yukawa coupling.  As discussed in \Sec{sec:top_yukawa} and confirmed by explicit calculation, the top quark Yukawa coupling does not generate a quadratic or logarithmic divergence.  The finite correction to the Higgs mass parameter is given by
\begin{equation}
- \frac{3 f^2}{16 \pi^2} 9 
\frac{|y_1|^2 |y_2|^2 |y_3|^2  }{|y_2|^2 -|y_3|^2} 
\log \left( \frac{|y_1|^2 + |y_2|^2}{|y_1|^2 + |y_3|^2} \right) h_1^T h_1 \ ,
\label{eq:top_divergence}
\end{equation}
where the factor of $9$ mirrors the factor of $3$ in \Eq{eq:top_higgs}.  In addition, the top Yukawa generates new contributions to the Higgs quartic coupling $\lambda_{56}$, but they are loop suppressed and therefore insignificant compared to the tree-level quartics already present.  

In addition, one can show that the full potential generated by loops of the top sector preserves custodial $SU(2)$, despite the appearance of custodial violation in the $y_2$ term of \Eq{eq:top_yukawa}.  To understand this, first perform the field redefinition $\Sigma U^c \rightarrow \widetilde U^c$ and $Q^T S \Sigma S \Sigma^T \rightarrow \widetilde Q^T$, such that the non-derivative interactions take the form
\be
\mathcal{L}_t = y_1f \, \widetilde{Q}^T \, \widetilde{U}^c + y_2 f\, {Q'_a}^T \,  \widetilde{U}^c + y_3 f \, \widetilde{Q}^T \, \widetilde{\Sigma}P_5 \, {U'_5}^c\, + \ {\rm h.c.} \ ,
\ee
where $P_5$ is the projection matrix defined in \Eq{eq:P56}. $P_5$ is really redundant here; it has been inserted purely for emphasis. $\widetilde{\Sigma}$ is defined as
\be
\widetilde{\Sigma} = \Sigma \, S \,  \Sigma^T \, S \, \Sigma \ .
\ee
In this basis, we see that the radiatively generated potential potential can only depend on $\Sigma$ through the 6-vector $\widetilde{\Sigma}_{i5}$.   $\widetilde{\Sigma}_{55}$ and $\widetilde{\Sigma}_{65}$ are invariant under $SO(4)$ and therefore cannot contribute to custodial symmetry violation. However, custodial symmetry violation in the $y_2$ term may manifest itself in the potential through bilinears which contract the upper four components of the vector $\widetilde{\Sigma}_{i5}\equiv v_i, i=1...4$ in an $SO(4)$ violating way. Such contractions must preserve the $SU(2)_L$ subgroup of $SO(4)$ and can be written in the general form
\be
c_0\,   v^T  v + c_a \, v^T T^a_R v  \ .
\ee
where $T_R^a$ are the $SU(2)_R$ generators. However, because of the antisymmetry of the $T_R^a$ generators, all contractions except for the fully $SO(4)$ symmetric one vanish. Thus, the radiative Higgs potential from the top Yukawa sector preserves custodial symmetry.

Loops of all other quarks (and leptons) contribute quadratically divergent mass terms $\delta m_1^2$ for the Higgs of the form
\begin{equation}
-\frac{3 \Lambda^2}{16 \pi^2}\, |y|^2\, h_1^T h_1 \ .
\label{eq:light_divergence}
\end{equation}
However, since the light fermion Yukawa couplings are all very small, none of these contributions are numerically significant.

\section{Physical Higgs Bosons}
\label{sec:physhiggs}

The mass matrix of the Higgses mixes $h_1$ and $h_2$ but it does not mix different components of the Higgs quartets, so it breaks up into four $2\times2$ matrices, one for each component.  Furthermore, by the remaining custodial symmetry, the three mass matrices in directions that do not get a vev are identical.  If we ignore small quartics generated by loops, then the mass of the Higgses in the unbroken directions is
\be
M_{A^0}^2 = M_{H^\pm}^2 = m_1^2+m_2^2 = \frac{2\, B_\mu}{\sin 2\beta}-\lambda_0 v^2,
\label{eq:Amass}
\ee
where the parameters are defined in Eqs.~(\ref{eq:higgs_pot}),~(\ref{eq:electroweak_vev}), and~(\ref{eq:tanbeta}), and the $A^0$ and $H^\pm$ are the pseudoscalar and charged Higgses respectively.   The $h_1$ and $h_2$ content of these is controlled by $\tan \beta$.  Small quartics and electromagnetic corrections will break the degeneracy between these states, but this is not a large effect.  

There are two Higgses in the broken direction whose masses are given by 
\be
M^2_{h^0,H^0} = \frac{B_\mu}{\sin 2\beta} \mp
\sqrt{ \frac{B_\mu^2}{\sin^2 2\beta} - 2 \lambda_0 B_\mu \, v^2 \sin 2\beta  +
\lambda_0^2 \, v^4 \sin^2 2\beta },
\ee
with
\begin{equation}
h^0 = \cos \alpha \; h_1 + \sin \alpha \; h_2, \qquad 
H^0 = \cos \alpha \; h_2 - \sin \alpha \; h_1,
\end{equation}
\be
\tan \alpha = \frac{1}{B_\mu -\lambda_0 v^2 \sin 2\beta } 
\left( B_\mu \cot 2 \beta + \sqrt{\frac{B_\mu^2}{\sin^2 2 \beta} - 
2 \lambda_0 B_\mu \, v^2 \sin 2 \beta
+\lambda_0^2 \, v^4 \sin^2 2 \beta} \right).
\ee
The light $h^0$ will generally be lighter than the charged Higgs $H^\pm$, while the heavy $H^0$ is usually heavier.  Again, there will be small corrections to these relations from small quartics, but the overall structure is unchanged. In the limit where $\tan \beta$ is large, we have
\begin{equation}
M_{h^0}^2 < \frac{4 \lambda_0}{\tan^2 \beta} v^2 \ , 
\end{equation}
which would be below the experimental direct search bound, so this model predicts $\tan \beta \simeq 1$.  Furthermore, from the radiative corrections discussed in \Secs{sec:modulargauge}{sec:top_yukawa}, we expect $m_2 > m_1$ or $\tan \beta > 1$. 

To summarize the most important features of the spectrum: custodial symmetry predicts a triplet of approximately degenerate Higgs states $A^0$ and $H^\pm$ whose masses are given by the free parameter $m_1^2+m_2^2$. The masses of the two remaining Higgs states are controlled predominantly by two other free parameters, $\lambda_0$ for the light Higgs and $B_\mu$ for the heavy Higgs. Thus the Higgs masses can vary over a relatively large range subject only to the custodial symmetry relation and experimental constraints.

\section{Alternative Fermion Charges}

\subsection{Top Sector}
\label{sec:alttop}

The phenomenology of a little Higgs theory depends strongly on the details of the top sector.  In this appendix, we mention two alternative mechanisms for generating a collective top Yukawa coupling compared to the preferred method detailed in \Sec{sec:top_yukawa}.

In the text, we implemented a collective top Yukawa coupling by taking $Q$ and $Q'_a$ to transform as multiplets of $SO(6)_A$ and $U^c$ and $U'^c_5$ to transform as multiplets of $SO(6)_B$.  Since $\Sigma$ is a bifundamental of $SO(6)_A \times SO(6)_B$, this determines the allowed form for the top interactions
\be
\label{eq:option3langrange}
\mathcal{L}^{(3)}_t = y_1f \, Q^T S \, \Sigma\,  S \, U^c + y_2 f\, {Q'_a}^T \, \Sigma \, U^c
+ y_3 f \, Q^T \, \Sigma \, U'^c_5\, + \ {\rm h.c.} \ .
\ee
Here, $S={\rm diag}(1,1,1,1,-1,-1)$ is an $SO(6)$ matrix necessary to make sure that all of the symmetries protecting the Higgs mass are broken by the top sector, and the $(3)$ superscript notation will be evident below.  The heavy top partners have masses
\be
\label{eq:option3masses}
|y_1| f, \qquad \sqrt{|y_1|^2 + |y_2|^2} f, \qquad  \sqrt{|y_1|^2 + |y_3|^2} f , 
\ee
and the low energy top Yukawa coupling is
\be
\label{eq:option3top}
y^{(3)}_t = 3 y_{123}, \qquad y_{123} \equiv \frac{y_1 y_2 y_3 }{\sqrt{|y_1|^2 + |y_2|^2} \sqrt{|y_1|^2 + |y_3|^2}} \ . 
\ee

The reason the factor of 3 appears in \Eq{eq:option3top} (and the reason for the $(3)$ superscript) is that the field $\Sigma$ appears in all three of the terms in \Eq{eq:option3langrange}.  That is, once the fermion masses are diagonalized, the $y_1$, $y_2$, and $y_3$ terms each contain couplings of the SM fermions to the Higgs modes.  For fixed values of $y_t$ and $f$, this allows the top partners to be a factor of 3 lighter than in a naive implementation of the top Yukawa where $y_t = y_{123}$ (see \Eq{eq:option1langrange} below).  Given the quadratic sensitivity of the Higgs boson mass to $m_T$, the fine-tuning can be 9 times less severe with this method of generating the top Yukawa.

In alternative top sectors, $\Sigma$ can appear in either one or two terms, leading to $\mathcal{L}^{(1)}_t$ and $\mathcal{L}^{(2)}_t$.  For reasons to be discussed below, having more than 3 insertions of $\Sigma$ is not radiatively stable.  To see why only one $\Sigma$ insertion is necessary, consider the case where $Q$ and $U'^{c}_5$ are $\mathbf{6}$'s of $SO(6)_A$ while $U^c$ and $Q'_a$ are $\mathbf{6}$'s of $SO(6)_B$.  A viable top sector is
\be
\label{eq:option1langrange}
\mathcal{L}^{(1)}_t = y_1f \, Q^T \, \Sigma \, U^c + y_2 f\, {Q'_a}^T  \, U^c
+ y_3 f \, Q^T  \, {U'_5}^c\, + \ {\rm h.c.} \ .
\ee
This type of top sector appears in many little Higgs constructions, e.g.\ \Ref{Katz:2003sn}.  Here, the top partners have the same masses as \Eq{eq:option3masses}, but the low energy top Yukawa coupling would be
\be
y^{(1)}_t = y_{123} \ .
\ee
Thus, for fixed $y_t$ and fixed ratios of the $y_i$, the top partners would be a factor of $3$ heavier compared to $\mathcal{L}^{(3)}_t$, drastically decreasing the observability of the top partners at colliders and increasing the fine-tuning in the top sector by a factor of $9$.  One important change in the phenomenology is that now the mass eigenstate and interaction eigenstate bases are twisted, such that the heavy top partners can decay to SM top and bottom quarks through emissions of the $\phi$/$\eta$ states.

An intermediate possibility is to arrange $\Sigma$ to appear in two of the terms in the collective top sector.  For example, let $Q$ and $U^c$ be $\mathbf{6}$'s of $SO(6)_A$ and let $Q'_a$ and $U'^c_5$ be $\mathbf{6}$'s of $SO(6)_B$.  The top sector could then be
\be
\label{eq:option2langrange}
\mathcal{L}^{(2)}_t = y_1f \, Q^T \,  S \, U^c + y_2 f\, {Q'_a}^T \,  \Sigma^T \, U^c
+ y_3 f \, Q^T \, \Sigma  \, {U'_5}^c\, + \ {\rm h.c.} \ ,
\ee
where $S$ is necessary in the first term to break all of the global symmetries protecting the Higgs mass.  This yields the same top partner masses as \Eq{eq:option3masses}, but an intermediate value for the top Yukawa coupling
\be
y^{(2)}_t = 2 y_{123} \ .
\ee
Thus for fixed $y_t$, the top partners are a factor of $3/2$ heavier than for $\mathcal{L}^{(3)}_t$, and the tuning worse by $9/4$.

With the observation that $\mathcal{L}^{(3)}_t$ gives a more favorable phenomenology compared to $\mathcal{L}^{(1)}_t$ and $\mathcal{L}^{(2)}_t$, one might wonder if it is possible to have a top sector $\mathcal{L}^{(n)}_t$ which involves $n$ insertions of $\Sigma$.  For example, consider a case which has the same fermion charge assignments as \Eq{eq:option1langrange}:
\be
\label{eq:optionnlangrange}
\mathcal{L}^{(2m+1)}_t = y_1f \, Q^T \, \Sigma \,  (S \, \Sigma^T \, S \, \Sigma)^m  \, U^c + y_2 f\, {Q'_a}^T  \, U^c
+ y_3 f \, Q^T  \, {U'_5}^c\, + \ {\rm h.c.} \ .
\ee
Naively, this model would yield a top Yukawa of $(2m+1) y_{123}$.  However, the form of \Eq{eq:optionnlangrange} is not radiatively stable, as $\Sigma$ loops will generate the operator $Q^T \, \Sigma \, U^c$, whose coefficient will be enhanced by combinatoric factors, making it dominate over the original $y_1$ term.  Thus,  $\mathcal{L}^{(2m+1)}_t$ essentially reduces to $\mathcal{L}^{(1)}_t$, which is not surprising given that they share the same fermion charge assignments.  One can easily convince oneself that given the fields $Q$, $Q'_a$, $U^c$, and $U'^c$, at most three $\Sigma$ insertions are radiatively stable.

\subsection{Leptons and Precision Electroweak}
\label{sec:new-leptons}

\begin{figure}
\includegraphics[scale=0.6]{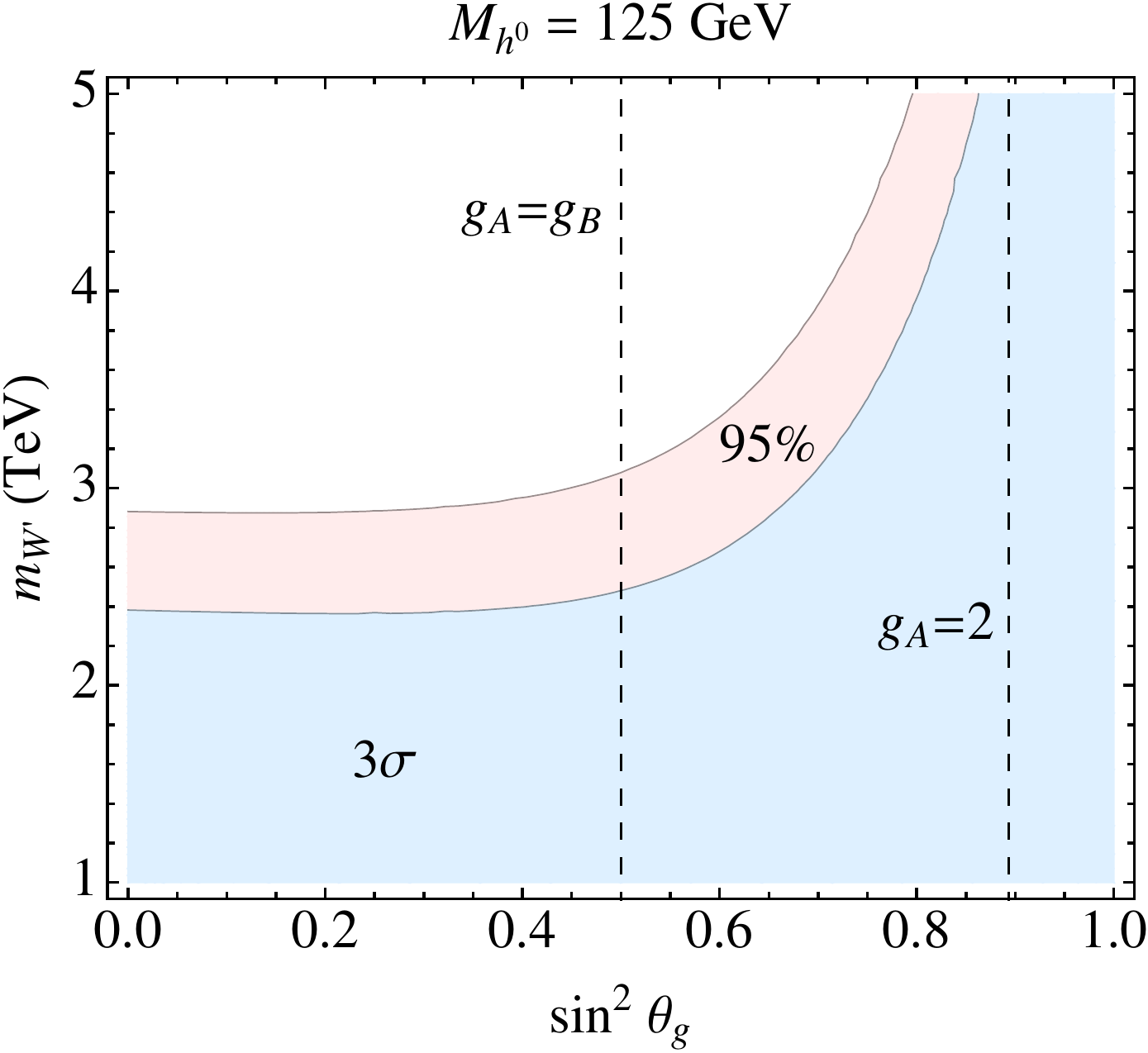}
\includegraphics[scale=0.6]{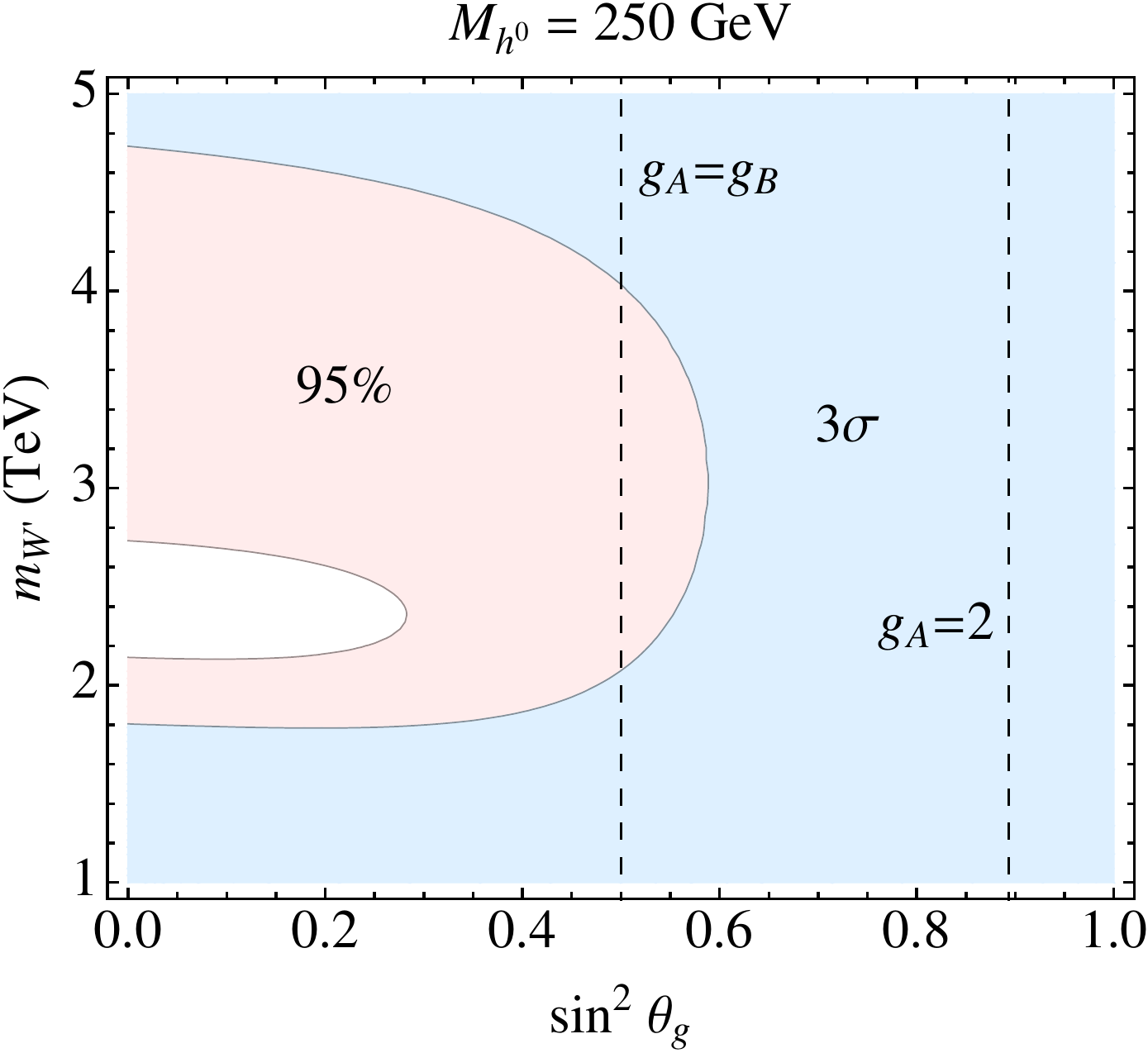}
\caption{Same as \Fig{fig:electroweak-fit} except with the lepton doublet charged under $SU(2)_A$ instead of $SU(2)_B$.}
\label{fig:other-electroweak-fit}
\end{figure}

In \Sec{sec:light_quarks}, we noted that we have a choice as to whether the SM lepton doublets should be charged under $SO(6)_A$ or $SO(6)_B$.  While the bestest model has the lepton doublet charged under $SO(6)_B$, in this appendix we explore how the electroweak fits change if the leptons couple the same way as the quarks.   With this assignment, we can calculate the coefficients of the dimension six operators:
\begin{align}
a^t_{ql} &= a^t_{ll} = - \frac{ g_A^4}{4 m_{W'}^2(g_A^2 + g_B^2)} = 
      - \frac{ g_{\rm EW}^2}{4 m_{W'}^2}\tan^2 \theta_g \, ,  \nonumber\\ 
a^t_{hq} &= a^t_{hl} =  -\frac{ g_A^2\, (g_A^2 - g_B^2) }{4 m_{W'}^2 (g_A^2 + g_B^2)} = 
      \frac{g_{\rm EW}^2}{4 m_{W'}^2}(1-\tan^2\theta_g)   \,     .
\end{align}
Since the leptons now have the same couplings to the heavy gauge bosons as the quarks which is shown in \Eq{eq:quark_Wp}, the $a$ coefficients are symmetric under interchange of $q$ and $l$. We repeat the fit to precision electroweak data using \cite{Han:2004az}. Our results for two different Higgs masses $M_{h^0} = 125$ and $250$ GeV are shown in \Fig{fig:other-electroweak-fit}.

Because the fermion couplings to the heavy gauge bosons are all proportional to $g_A$, the region of large $g_A$ (equivalent to large $\sin^2 \theta_g$) has much more stringent bounds than the region of small $g_A$.  For $M_{h^0} = 125$ GeV, there is a large region of allowed parameter space, but the mass of the gauge partners is always $\gtrsim 3$ TeV.  On the other hand, for the heavier Higgs there is only a small region of parameter space allowed, though this region does have the gauge partner mass between 2 and 3 TeV.  Also, unlike in the model of \Sec{sec:light_quarks}, in no region is the fit improved with respect to the SM.  While this charge assignment is allowed, by comparing \Figs{fig:electroweak-fit}{fig:other-electroweak-fit}, we conclude that the model presented in \Sec{sec:light_quarks} is indeed the bestest.

\end{document}